\documentclass[sigconf, nonacm]{acmart}
\makeatletter

\def\@ACM@checkaffil{
    \if@ACM@instpresent\else
    \ClassWarningNoLine{\@classname}{No institution present for an affiliation}%
    \fi
    \if@ACM@citypresent\else
    \ClassWarningNoLine{\@classname}{No city present for an affiliation}%
    \fi
    \if@ACM@countrypresent\else
        \ClassWarningNoLine{\@classname}{No country present for an affiliation}%
    \fi
}

\newcommand\addauthornote[1]{%
  \if@ACM@anonymous\else
    \g@addto@macro\addresses{\@addauthornotemark{#1}}%
  \fi}

\newcommand\@addauthornotemark[1]{\let\@tmpcnta\c@footnote
   \setcounter{footnote}{#1}\addtocounter{footnote}{-1}
    \g@addto@macro\@currentauthors{\footnotemark\relax\let\c@footnote\@tmpcnta}}

\makeatother

\usepackage{tikz}
\usetikzlibrary{shapes,arrows}
\usepackage{amsmath}

\usepackage[linesnumbered, vlined, noend, ruled]{algorithm2e}

\usepackage{amsmath}
\usepackage{amsfonts}
\usepackage{amsthm}
\usepackage{array}
\usepackage{cancel}
\usepackage{color}
\usepackage{colortbl}
\usepackage{courier}
\usepackage{comment}
\usepackage{framed}
\usepackage[T1]{fontenc}
\usepackage{hhline}
\usepackage{lipsum}
\usepackage{listings}
\usepackage{mathtools}
\usepackage{multicol}
\usepackage{multirow}
\usepackage{pifont}
\usepackage{soul}
\usepackage{setspace}
\usepackage{subcaption}
\usepackage{tabularx}
\usepackage{textcomp}
\usepackage{upquote}
\usepackage{hyperref}
\usepackage{varwidth}
\usepackage{xspace}
\usepackage{xcolor}
\usepackage{booktabs}
\usepackage{wasysym}

\usepackage{cleveref}

\usepackage{caption}
\AtBeginDocument{%
  \providecommand\BibTeX{{%
    \normalfont B\kern-0.5em{\scshape i\kern-0.25em b}\kern-0.8em\TeX}}}

\acmConference[]{}{}{}

\acmPrice{}
\acmISBN{}

\newcommand{\secref}[1]{\mbox{Section~\ref{#1}}}

\newcommand{\fref}[1]{\mbox{Fig.~\ref{#1}}}
\newcommand{\tabref}[1]{\mbox{Table~\ref{#1}}}

\newcommand{\PP}[1]{
\noindent{\bf \IfEndWith{#1}{.}{#1}{#1.}}
}
\newcommand{\PPP}[1]{
\noindent{\bf \IfEndWith{#1}{}{#1}{#1}}
}
\newcommand{\PPPP}[1]{
\indent{\bf \IfEndWith{#1}{}{#1}{#1}}
}
\newcommand{\PPPPP}[1]{
\indent{\bf \IfEndWith{#1}{.}{#1}{#1.}}
}

\newcommand*\circled[1]{\tikz[baseline=(char.base)]{
            \node[shape=circle,fill,inner sep=1pt] (char) {\textcolor{white}{#1}};}}

\usepackage{hyperref}
\definecolor{linkcolour}{rgb}{0,0.2,0.6}
\definecolor{xgreen}{rgb}{0.2,0.6,0.0}
\definecolor{xred}{rgb}{0.7,0.1,0.0}
\definecolor{Gray}{gray}{0.92}
\hypersetup{colorlinks,breaklinks,citecolor=xred, urlcolor=linkcolour, linkcolor=xgreen}

\newcommand{\sys}{\mbox{\textsc{Bridge}}\xspace }

\newcommand{\scadaphysics}{\textsc{scada-physics}\xspace}

\newcommand{\wwrite}{\textsc{write}\xspace}
\newcommand{\rread}{\textsc{read}\xspace}

\newcommand{\feng}{\mbox{\textsc{Invariant}}\xspace }
\newcommand{\scaphy}{\mbox{\textsc{Scaphy}}\xspace }
\newcommand{\scada}{\mbox{\textsc{Scada}}\xspace }

\newcommand{\accuracy}{{98.3\% }}

\newcommand{\attacks}{{\small{50 }}}
\newcommand{\existingaccuracy}{{78.3\% }}

\newcommand{\scenarios}{\mbox{\textsc{11}}\xspace }

\newcommand{\fp}{{0.8\% }}
\newcommand{\existingfp}{{13.7\% }}

\newcommand{\rd}{\mbox{\textsc{$R_{D}$}}\xspace }
\newcommand{\sv}{\mbox{\textsc{$S_V$}}\xspace }
\newcommand{\cv}{\mbox{\textsc{$C_V$}}\xspace }

\newcommand{\rdt}{\mbox{\textsc{$R_{D\Delta}$}}\xspace }
\newcommand{\rdf}{\mbox{\textsc{$R_{D\digamma}$}}\xspace }
\newcommand{\rdb}{\mbox{\textsc{$R_{D\mu}$}}\xspace }

\newcommand{\fillv}{\emph{Intake Valve}\xspace }
\newcommand{\dosingv}{\emph{Supply Valve}\xspace }

\newcommand{\intakev}{\emph{Intake Valve}\xspace }
\newcommand{\supplyv}{\emph{Supply Valve}\xspace }

\usepackage{xstring}

\makeatletter
\newcommand{\labitem}[2]{%
    \def\@itemlabel{\textbf{#1}}
	\item
	\def\@currentlabel{\textbf{#1}}\label{#2}}
\makeatother

\SetKwProg{Fn}{Func}{}{}

\usepackage{makecell,rotating} 
\begin{document}

\title[\sys: Bridging Both Worlds in Semantics and Time]{Bridging Both Worlds in Semantics and Time: Domain Knowledge Based Analysis and Correlation of Industrial Process Attacks}



\author{Moses Ike}
\affiliation{%
  \institution{Sandia National Laboratories}
}

\author{Kandy Phan}
\affiliation{%
  \institution{Sandia National Laboratories}
}

\author{Anwesh Badapanda}
\affiliation{%
  \institution{Indian Institute of Technology Delhi}}

\author{Matthew Landen}
\affiliation{%
  \institution{Georgia Institute of Technology}
}

\author{Keaton Sadoski}
\affiliation{%
  \institution{Sandia National Laboratories}
}

\author{Wanda Guo}
\affiliation{%
  \institution{Texas A\&M University}}

\author{Asfahan Shah}
\affiliation{%
  \institution{Bennett University}
}

\author{Saman Zonouz}
\affiliation{%
  \institution{Georgia Institute of Technology}}

\author{Wenke Lee}
\affiliation{%
  \institution{Georgia Institute of Technology}
}



\begin{abstract}
Modern industrial control systems (ICS) attacks infect supervisory control and data acquisition (\scada) hosts to \emph{stealthily} alter industrial processes, causing damage. To detect attacks with low false alarms, recent work detects attacks in both \scada and process data. Unfortunately, this led to the same problem - disjointed (false) alerts, due to the semantic and time gap in \scada  and process behavior, i.e., \scada execution does not map to process dynamics nor evolve at similar time scales. We propose \sys to analyze and correlate \scada and industrial process attacks using domain knowledge to bridge their unique semantic and time evolution. 
This enables operators to tie malicious \scada operations to their adverse process effects, which reduces false alarms and improves attack understanding. \sys (i) identifies process constraints violations  in \scada by measuring actuation dependencies in \scada process-control, and (ii) detects malicious \scada effects in processes via a physics-informed neural network that embeds generic knowledge of \emph{inertial process dynamics}.  \sys then dynamically \emph{aligns} both analysis (\emph{i} and \emph{ii}) in a time-window that \emph{adjusts} their time evolution based on process inertial delays.
We applied \sys to \scenarios diverse real-world industrial processes, and adaptive attacks inspired by past events. \sys correlated \accuracy of attacks with \fp false positives (FP), compared to \existingaccuracy detection accuracy and \existingfp FP of recent work.

\end{abstract}



\keywords{industrial processes, industrial control system, SCADA}
\settopmatter{printfolios=true}

\maketitle

\section{Introduction}
\label{sec:introduction}
Industrial control systems (ICS) operate life-essential industrial processes such as water treatment and manufacturing plants. In ICS networks, supervisory control and data acquisition (\scada) hosts manage processes, comprised of actuators, sensors, and programmable logic controllers (PLCs). Processes are governed by the dynamics (or physics) of their actuators.
\scada runs software and Human Machine Interfaces (HMI) to control processes (called \emph{process-control}~\cite{scaphy}). Unfortunately, modern attacks (e.g., Industroyer, Stuxnet~\cite{industroyer, stuxnet}) infect \scada to disrupt processes~\cite{attacksok2,attacksok3,attacksok4,risi-database,icsstrive}. For example, the 2021 Oldsmar water treatment attack infected a \scada HMI to alter process parameters and poison the water supply~\cite{scaphy,florida,florida2,oldsmartech,attacksok}.
 
To accurately detect industrial process attacks, it is essential to connect the \emph{causal} software action in \scada to the adverse physical \emph{effect} on the process. On one hand, process anomaly detection~\cite{tabor, limiting, state} such as \feng~\cite{invariants} can detect abnormal sensor data, but is evaded by small/\emph{stealthy} process deviations by \scada adversaries, which over time causes damages~\cite{incrementattack, timedelay, timedelay2}. On the other hand, host anomaly detection~\cite{orpheus, lee} cannot know if abnormal \scada execution (e.g., system calls) has adverse process effects. Because these (disjointed) techniques cannot tie the \scada \emph{cause} and process \emph{effect} of attacks, they cause high false alarms due to benign faults and errors~\cite{scaphy}. Further, PLC defenses (e.g., VetPLC~\cite{vetplc}) can detect altered PLC logic but are not suited for \scada adversaries, who can alter PLC parameters at run-time without touching its logic. 

Identifying malicious \scada actions \emph{that cause} adverse process effects, and tying them \emph{in time} to the process behavior is challenging because \scada and industrial processes differ in \emph{semantics} and \emph{time} evolution.  
That is, while \scada runs \emph{discrete} software at CPU speed, processes are governed by \emph{continuous} physics laws, which constrain their actuation (e.g., \emph{inertial} delays). 
A recent work, \scaphy~\cite{scaphy}, analyzed both \scada and process data to detect attacks. \scaphy used Open Platform Communication (OPC) events~\cite{ae,ae1, opc, opc2, opc3} to induce system call signatures unique to \scada phases and physical states. However, its analysis in each domain was disjointed and not connected in semantics or time. This led to high false alarms from isolated alerts. In addition, \scaphy's reliance on signatures limits it to only known attacks, i.e., it cannot detect unknown attacks. As demonstrated in our experiments, attacks that use normal \scada tools/system calls (e.g., Oldsmar) will evade \scaphy's signatures.

Connecting \scada and processes semantically involves understanding process constraints in \scada operation, and alternatively, identifying \scada effects in process behavior. To control processes, \scada adheres to physics dependencies intrinsic to process tasks i.e., how actuators \emph{depend on} each together to achieve a task~\cite{significance}. 
Therefore, an ideal semantic relationship can be formulated from this "process adherence" in software.
However, \emph{discrete} system calls (used in \scaphy) cannot capture process dynamics, which are \emph{continuous}. At a high level, we first inspect relevant program-flow dependencies in process-control logic. For example, if \scada \emph{monitors} device A to \emph{actuate} device B, then B depends on A. Then, we statistically measure the \emph{continuous} behavior of these dependencies in process-control executions. Through this, an attacker's violations of intrinsic process constraints in \scada, will result in adverse process effects, which can now be selectively flagged in \scada.

To detect adverse process effects (i.e., due to \scada attack), neural networks such as autoencoders (AE)~\cite{rnn, rnn2} can learn normal process behavior via sensor-actuator time-series data. An AE aims to reconstruct its input as its output. Since there is compression in the AE inner layers, it is forced to learn relationships in the input distribution. This way, AEs trained on benign time series will have \emph{errors} reconstructing anomalous sequences (via a loss function), allowing adverse change effects to be detected. However, training data are limited in practice, which can limit AE's accuracy~\cite{physics-cps, aesecurity}. 

Further, processes experience \emph{inertial} forces (due to physics factors e.g., friction/momentum~\cite{inertia}) which \emph{resists change} in actuators. This introduces irregularities/noise to process dynamics~\cite{inertia,inertia2, inertia3}. This not only impairs AE's correctness but impedes the ability to analyze malicious \scada effects on processes, especially stealthy perturbations that blend with noise~\cite{incrementattack, timedelay, timedelay2}.

In addition, connecting \scada and processes in time aims to determine \emph{where} or \emph{when} a \scada attack can be noticeably detected in process behaviors. This is critical to \emph{tie} both attacks \emph{in time}.  
Recall that inertial forces can maintain process behavior for a while even when under attack. For example, a conveyor belt actuated by large discs will not stop/slow down at once after a \emph{stop} command. As such, when \scada attacks are flagged, the adverse effect may not be present in the same time window, which leads to false negatives (when correlated), but \emph{later} causes damage. 
Further, because AEs analyze process inputs \emph{sequentially}, this causes more time discrepancies in correlating with the faster \scada side. 
We present \sys, a domain knowledge-based analysis and correlation of \scada and industrial process attacks that bridges their unique semantic and time evolution. This enables operators to tie malicious \scada execution to their adverse process effects, which reduces false alarms and improves attack understanding. At the \scada side, \sys analyzes the physical dependency constraints of \scada process-control and measures their statistical behavior such as control frequencies and bursts. 
Because process-control behavior \emph{vary} based on a process setpoint (e.g., what level to fill a tank), \sys develops a statistical solution based on \emph{coefficient of variations}~\cite{cv, cv2}, which aggregates measured constraints to work for processes regardless of their calibrated setpoints. 
Using these constraints, \sys selectively monitors and flags a \scada attacker's executions that violate intrinsic process semantics. 

On the process side, \sys develops a \emph{physics-informed} neural network (PINN)~\cite{pinn1, pinn2} that embeds physical laws on \emph{inertial process dynamics} via a partial differential equation (PDE) loss function. This
facilitates it to capture the right solutions and generalize well even in limited data. 
To mitigate sequential delays in process time-series, \sys implements the PINN in a Transformer-based architecture, which instantiates parallel \emph{Attention} processes~\cite{attention} for each sequence input, enabling timely correlation with \scada.
\begin{figure}[t]
\centering
\includegraphics[width=0.43\textwidth]{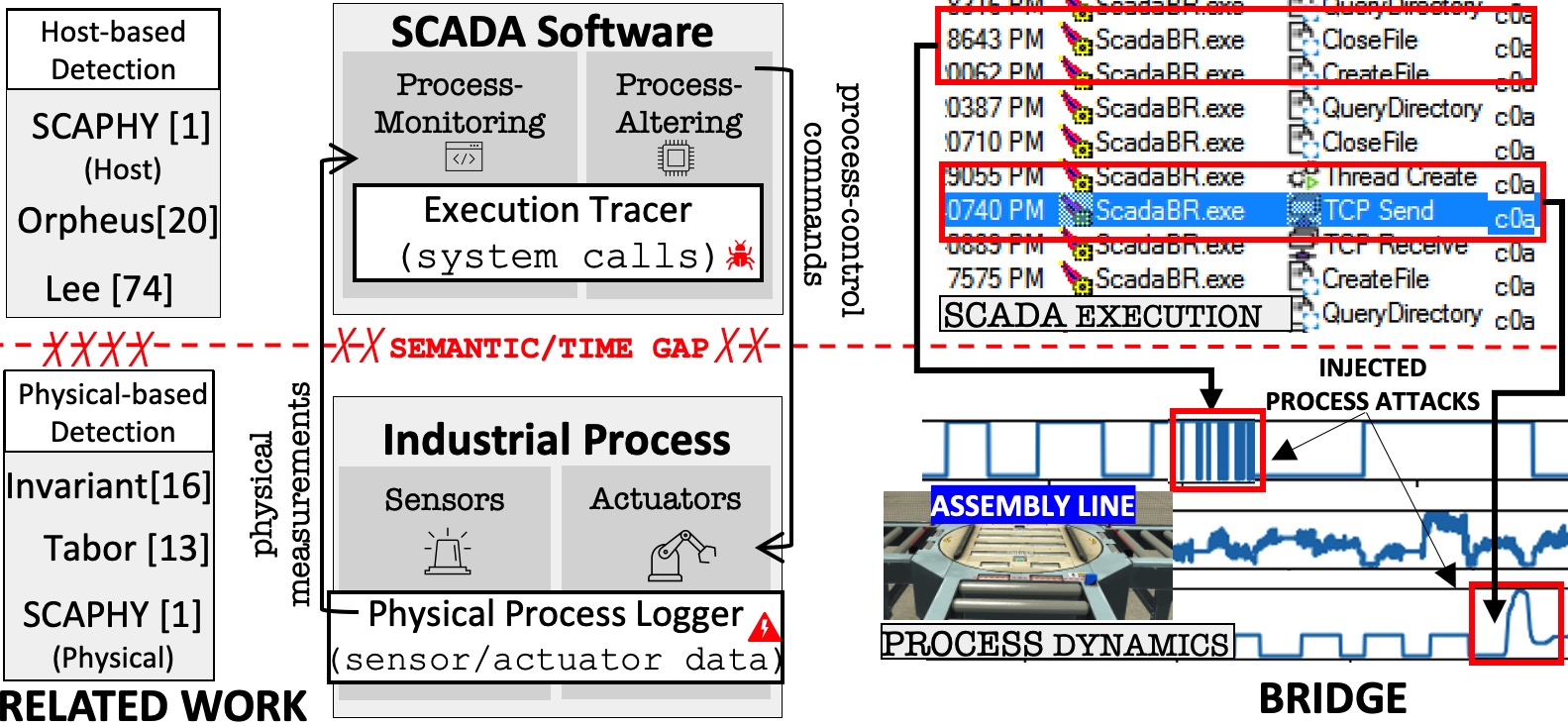}
  \caption{Showing how \sys ties malicious \scada operation to adverse process effects in contrast to existing (disjointed) techniques}
\label{fig:connect}
\end{figure}

To align \scada and process evolution in time (to know where \scada attacks can be noticeably detected in processes), \sys first uses a hybrid data-driven approach to derive the inertia delay associated with a process task, called \emph{inertia time-block} (ITB). It does this by analyzing the \emph{rate of change} of relevant process values. After a \scada attack is flagged, \sys waits for one ITB before checking for process anomalies (i.e., process effects lag behind \scada by at least 1 ITB). Then, it continues to analyze anomalies using a dynamic process evolution window, which interleaves multiple ITBs until the process rate of change reaches a steady state (i.e., no change). Through this, \sys derives an evolution window that \emph{bounds} the temporal lifespan of \scada effects on processes.  

\sys's correlation approach (shown in \fref{fig:connect}) detects attacks that current work miss such as attacks that infect benign \scada tools to \emph{stealthily alter} processes (e.g., Oldsmar attack), but over time cause damages. In these stealthy attacks, recent tools such as \feng and \scaphy will either discard the small deviation as false alarms or (if they narrow their thresholds), will trigger other benign deviations as attacks. This is because additional evidence (i.e., \sys's semantic connection of a prior \scada anomaly in the process evolution window) is needed to \emph{confirm} the attack and discard false alarms by current (disjointed) analysis.
Since modern attacks~\cite{attacksok2, attacksok3, attacksok4, risi-database, icsstrive} infects \scada to attack processes, we leverage this real-world ICS threat-model to apply process anomalies as a \emph{filter} for \scada attacks. That is, although process anomalies are analyzed, they are not used unless a \scada attack is flagged in the process evolution, i.e., the effect cannot happen before the cause.  

Since \sys uses generic \scada and sensor/actuator data, it is \emph{device-agnostic} and applies to industrial processes based on the widely-deployed ICS Purdue model~\cite{purdue}. Further, since inertia is a physics property of actuation systems, our PINN is generic to ICS processes but vital for correlation with \scada. Unlike \scaphy, which makes attack assumptions using signatures, \sys \emph{generalizes} attacks by detecting anomalies that \emph{violates} intrinsic semantic constraints in industrial processes. We make these contributions:
\begin{figure*}[t]
\centering
\includegraphics[width=0.69\textwidth]{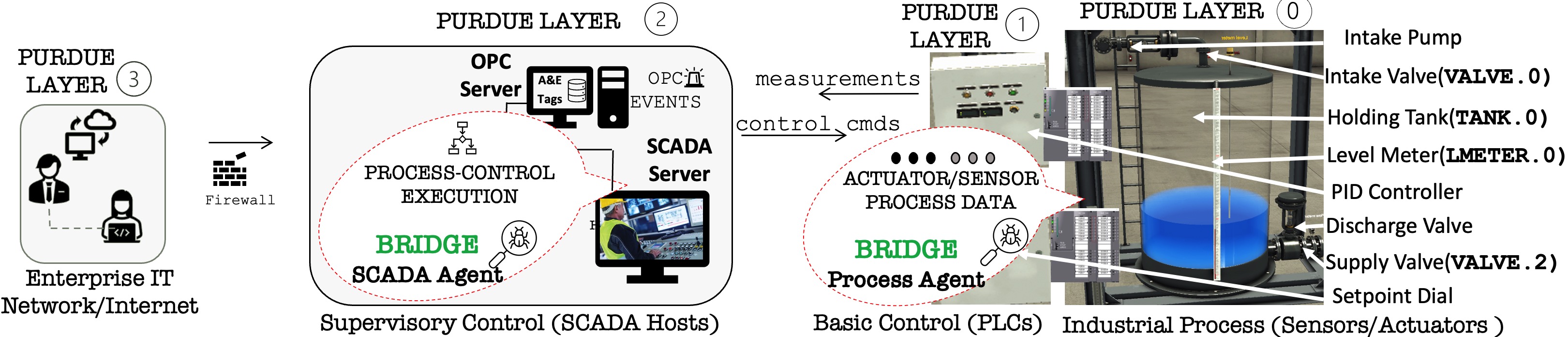}
  \caption{Showing where \sys is applied in real-world ICS environments, i.e., to analyze the \scada and process logs of the whole ICS}
\label{fig:architecture}
\end{figure*}
\begin{figure*}[t]
\centering
\minipage{0.21\textwidth}%
\begin{subfigure}[c]{\textwidth}
  \includegraphics[width=0.99\textwidth]{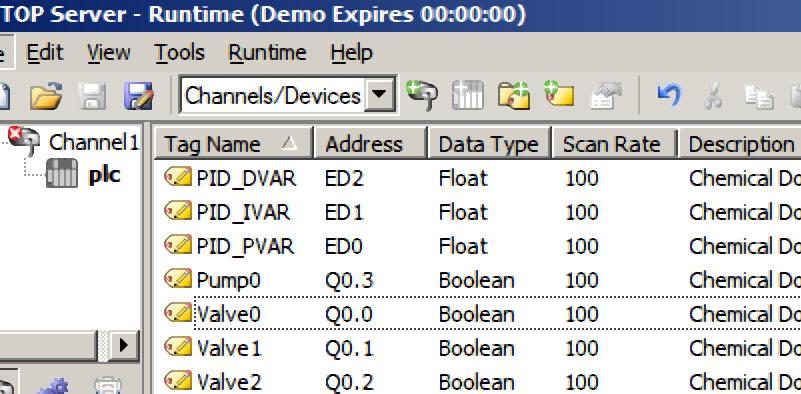}
\caption{OPC Device Tags}
\label{fig:opc-devices}
\end{subfigure}
\endminipage\hspace{.3em}%
\minipage{0.19\textwidth}%
\begin{subfigure}[c]{\textwidth}
  \includegraphics[width=0.98\textwidth]{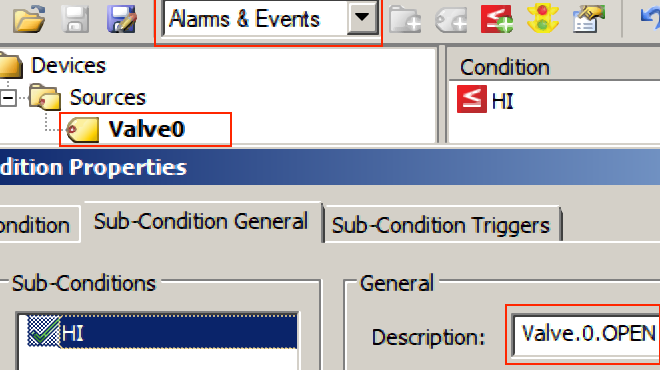}
\caption{OPC Event Config}
\label{fig:opc-ae}
\end{subfigure}
\endminipage\hspace{.3em}%
\minipage{0.12\textwidth}%
\begin{subfigure}[c]{\textwidth}
\includegraphics[width=0.99\textwidth]{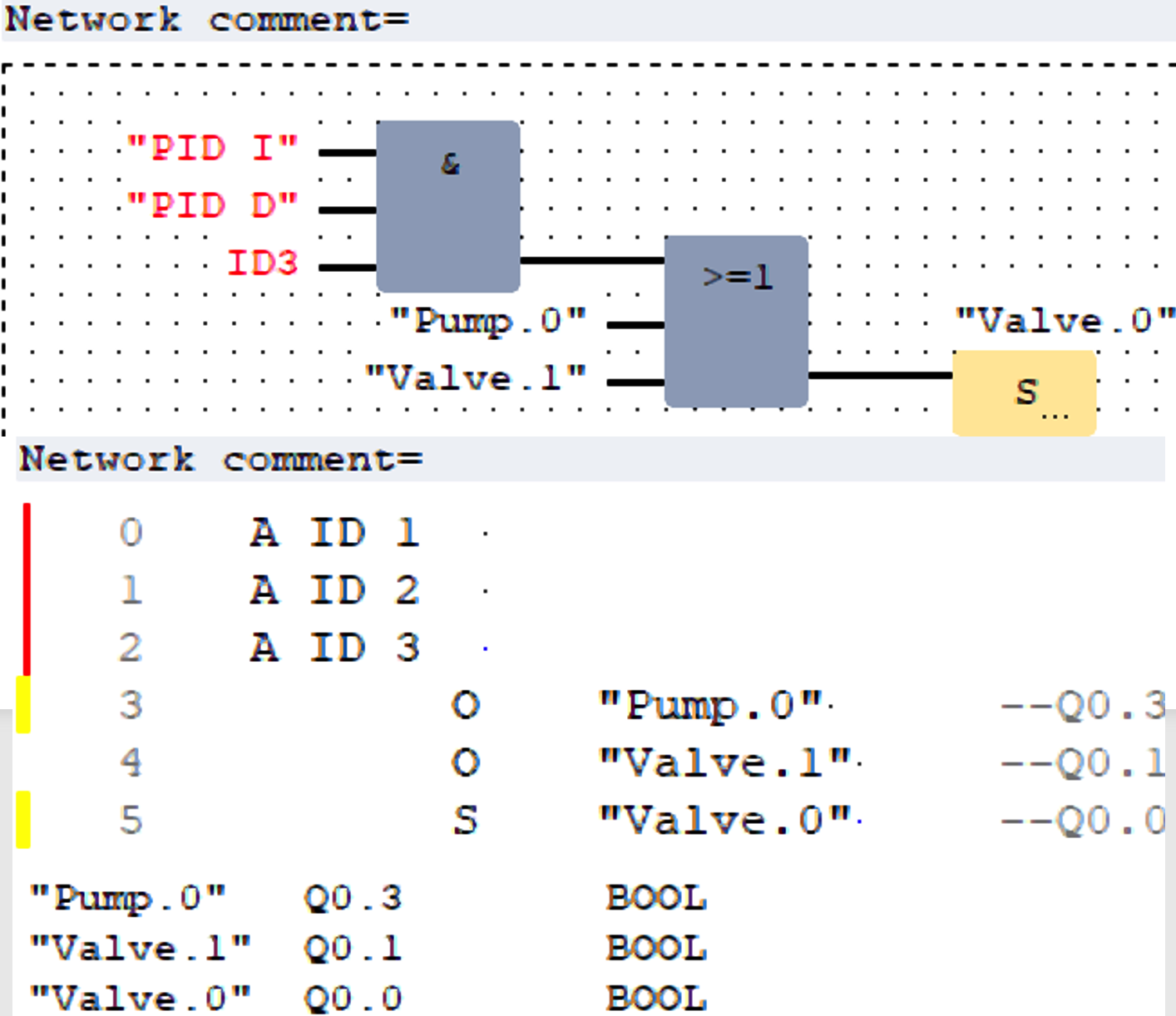}
  \caption{FBD and STL}
  \label{fig:stl}
 \end{subfigure}
\endminipage\hspace{.3em}%
\minipage{0.35\textwidth}%
\begin{subfigure}[c]{\textwidth}
\includegraphics[width=0.99\textwidth]{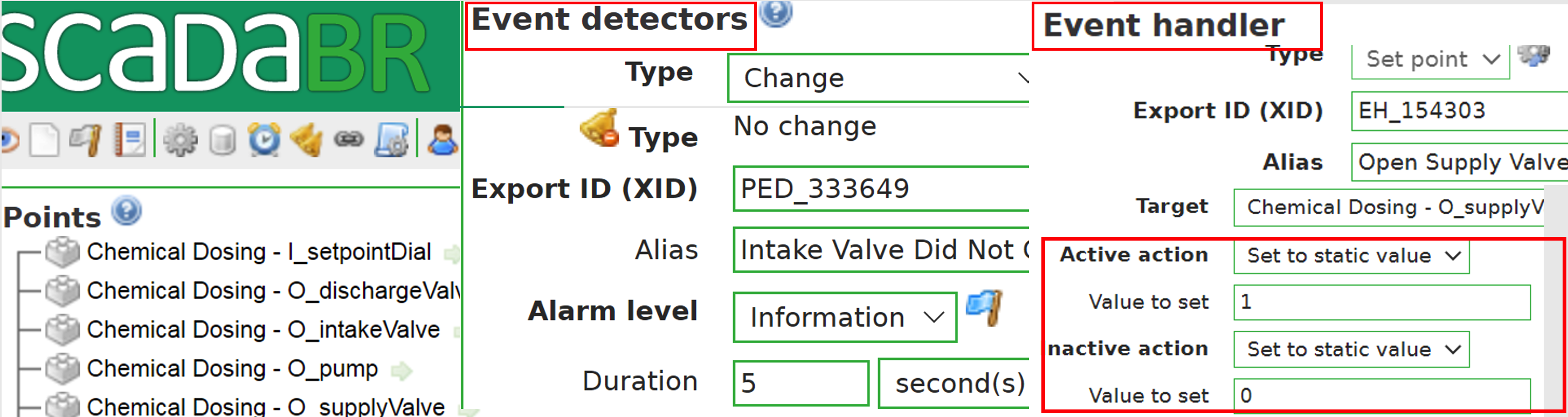}
  \caption{Industrial configurations of SCADA event-Handlers}
\label{fig:scadabr}
\end{subfigure}
\endminipage\hspace{.3em}%
\caption{Showing real-world \scada configurations of process sensors, actuators, and OPC events in Industrial Environments}
\label{fig:opc-setup}
\end{figure*}
\begin{enumerate}
  \item We present a new technique to correlate \scada and process attacks both semantically and time-wise via domain knowledge that bridges their unique behavior and time evolution. 
  \item We introduce a physics-informed learning architecture that captures inertial dynamism in processes via a domain PDE loss function deployed in a Transformer-based AE. This not only increases the learning robustness in limited data, but enables it to timely and accurately correlate with \scada. 
    \item We propose a new technique to derive intrinsic process constraints from \scada execution, which works for unique processes regardless of their setpoint calibration.
  \item We performed extensive experiments and case studies to validate \sys using public real-world data from \scenarios diverse industrial processes\footnote{https://github.com/lordmoses/\scaphy} and attacks inspired by past events.
We compared \sys to \scaphy~\cite{scaphy} and \feng~\cite{invariants}. \sys correlated \accuracy of attacks with \fp false positives (FP), compared to \existingaccuracy detection and \existingfp FP in the second best work, \scaphy. \sys is available online\footnote{https://anonymous.4open.science/r/bridge/}.
\end{enumerate}

\section{Background And Motivation}
In this section, we describe ICS operations based on the Purdue network model. To motivate our problem, we use a real-world attack running example to show how \sys is applied in the \scada and process sides. We then analyze the challenges of detecting modern attacks and existing work limitations in comparison to \sys.

\subsection{Industrial Processes and Control Operations}
Physical tasks in industrial plants are called processes~\cite{scaphy}. Without loss of generality, \fref{fig:architecture} describes the \scada and physical domain of ICS using a \emph{chemical dosing} process in a real-world water treatment plant. Actuators (e.g., pumps, valves) at Purdue layer 0 work together to achieve the process task, such as filling a tank to a \emph{setpoint} level (\sv). Actuating systems are governed by physics laws such as inertia. For example, the opening (speed) of a solenoid valve depends on the strength of induced electromagnetic fields.  These dynamic physics factors constrain industrial process behaviors (called \emph{process dynamics}) and time evolution. 
Further, PLCs at layer 1 set the \emph{basic} inputs of actuators at every scan cycle. \scada hosts at layer 2 perform \emph{supervisory} control of all actuators and PLCs. They run special software to monitor/respond to process events and manage processes to achieve their task~\cite{significance, siemens-water}.\begin{table*}[t]
\caption{Taxonomy of representative related works in ICS defense categorized by their techniques and approach}
    \centering
    \footnotesize
        \resizebox{0.97\textwidth}{!}{
        \begin{tabular}{@{}l|llll|llllllll|     lll|llll        |lllllll|llllllll|ll|l|>{\columncolor[gray]{0.9}}l}
            \toprule
            &\multicolumn{12}{c|}{\textbf{ICS Traffic Analysis}}&\multicolumn{14}{c|}{\textbf{Physical Behavior Analysis}}&\multicolumn{8}{c|}{\textbf{PLC Logic Analysis}}&\multicolumn{4}{c}{\textbf{SCADA Host Exec.}}\\
            \cline{2-39} 
            &\multicolumn{4}{c|}{\textbf{Sequences}}  &\multicolumn{8}{c|}{} &\multicolumn{3}{c|}{\textbf{R.  Learning}}&\multicolumn{4}{c|}{\textbf{Process-Aware}}&\multicolumn{7}{c|}{}&\multicolumn{8}{c|}{}&\multicolumn{2}{c|}{}&\multicolumn{2}{c}{\textbf{+ Physical}}\\
            \cline{2-5} 
            \cline{14-20} 
            \cline{38-39}
            Techniques
            &\rotatebox{90}{Critical~\cite{criticalstate}} &\rotatebox{90}{State-IDS~\cite{pattern2}} &\rotatebox{90}{Pattern~\cite{pattern1}} &\rotatebox{90}{SPEAR~\cite{anotherml}} 
            &\rotatebox{90}{Yang~\cite{deeplearning}} &\rotatebox{90}{Ihab~\cite{bayes}}&\rotatebox{90}{Ponoma.~\cite{telemetry}} &\rotatebox{90}{Diversity.~\cite{bullet}}
            &\rotatebox{90}{Evolution.~\cite{evolution} }
            &\rotatebox{90}{ML PIDS ~\cite{Ayodeji}}&\rotatebox{90}{Experion ~\cite{Honeywell}}  &\rotatebox{90}{Blockchain ~\cite{blockchain}}  
             &\rotatebox{90}{Kurt~\cite{rl1}} &\rotatebox{90}{Zhong~\cite{rl2}} &\rotatebox{90}{Panfili~\cite{rl3}}
            &\rotatebox{90}{Chromik~\cite{bro}}
            &\rotatebox{90}{Nivethan~\cite{process-semantics}}
            &\rotatebox{90}{Remke~\cite{process-awareness}}
            &\rotatebox{90}{Lin~\cite{runtime-powerflow}}
            &\rotatebox{90}{Ghaeini~\cite{state}} &\rotatebox{90}{Dina~\cite{ar}} &\rotatebox{90}{Aoudi~\cite{ssa}} &\rotatebox{90}{OSCIDS ~\cite{Ontology}} &\rotatebox{90}{Abbasi ~\cite{memory}} &\rotatebox{90}{Invariant~\cite{invariants}} &\rotatebox{90}{TABOR~\cite{tabor}}
            &\rotatebox{90}{Niang~\cite{formal}} &\rotatebox{90}{Mulder~\cite{weaselboard}}&\rotatebox{90}{Formby~\cite{formby}}  &\rotatebox{90}{PLC-Sleuth~\cite{mitm3}} &\rotatebox{90}{PLCDefend~\cite{mitm2}} &\rotatebox{90}{NoisePrint~\cite{mitm4}}  &\rotatebox{90}{TSV~\cite{tsv}}&\rotatebox{90}{VETPLC~\cite{vetplc}}
            &\rotatebox{90}{Lee~\cite{lee}}  &\rotatebox{90}{Orpheus~\cite{orpheus}}  
            &\rotatebox{90}{\scaphy~\cite{scaphy}} &\rotatebox{90}{\sys} \\
            \cline{1-39}  
            Action Sequence &$\bullet$&&&&&&&&&&&&&&&&&&&&&&&&&&&&&&&&&&&$\bullet$&$\bullet$&$\bullet$\\
            State Transitions &$\bullet$&$\bullet$&$\bullet$&&&&&&&&$\bullet$&&&&&&&&&&&&&&&&&&&&&$\bullet$&&&&&&\\
            Event-based &&&&&&&&&&&&$\bullet$&&&&&&&&&&&&&&&&&&&&$\bullet$&&&&&$\bullet$&$\bullet$\\
           \cline{1-39}
            ICS Traffic &$\bullet$&$\bullet$&$\bullet$&$\bullet$&$\bullet$ &&$\bullet$&&$\bullet$&$\bullet$&$\bullet$&$\bullet$&&&&&&&&&&&&&&&&&&$\bullet$&$\bullet$&&&&&&& \\
            Response Analysis &&&&$\bullet$&$\bullet$ &$\bullet$&$\bullet$&&&&&&&&&&&&&&&&&&&&&&&&&&&&&&& \\
           \cline{1-39}
            Physics Modelling &&&&&&& &&&&&&&&&&&&&$\bullet$ &$\bullet$&$\bullet$&$\bullet$&&&$\bullet$&&&&&&&&&&&&$\bullet$  \\
           \cline{1-39}
        PLC Control Logic &&&&&&&&&&&&&&&&&&&&&&&&$\bullet$&&&$\bullet$&$\bullet$&$\bullet$&&$\bullet$&$\bullet$&$\bullet$&$\bullet$&&&&  \\
            Logic verification &&&&&&&&&&&$\bullet$&&&&&&&&&&&&&$\bullet$&&&&&$\bullet$&&&&$\bullet$&$\bullet$&&&&\\
            Control register &&&&&&&&&&&$\bullet$&$\bullet$&&&&&&&&&&&$\bullet$&&&&&&$\bullet$&$\bullet$&&&&&&&&\\ 
            \cline{1-39}

            Online POMDP &&&&&&&&& &&&&$\bullet$&& &&&& &&&&&&&&&&&&&&&&&&&\\
            Reward weights &&&&&&&&&&&& &&$\bullet$& &&&&&&&&&& &&&&&&&&&&&&&\\
            Multi-agent game &&&&&&&&&&&& &&$\bullet$&$\bullet$ &&&&&&&&&&&&&&&&&&&& &&&\\
            \cline{1-39}

            Rules/Signatures &&&&&&&&&&&& &&& &$\bullet$&$\bullet$&$\bullet$& &&&&&&&&&&&&&&&&&&$\bullet$&\\
            Power-flow rules &&&&&&&&&&&& &&& &&&$\bullet$&$\bullet$ & &&&&&&&&&&&&&&&&&&\\
          \cline{1-39}
             Host Execution&&&&&&&&$\bullet$&&&&&&&&&&&&&&&&&&&&&&&&&&&$\bullet$&$\bullet$&$\bullet$&$\bullet$ \\
             \cline{1-39}
             Semantic Correlate&&&&&&&&&&&&&&&&&&&&&&&&&&&&&&&&&&&&&x&$\bullet$ \\
             Time Correlation&&&&&&&&&&&&&&&&&&&&&&&&&&&&&&&&&&&&&x&$\bullet$ \\
           \bottomrule
        \end{tabular}
        }
    \label{tbl:taxonomy}
\end{table*} This is called \emph{process-control}. 
For example, Siemens SIMATIC water systems manage entire transition operations in water treatment plants~\cite{watersok, siemens-water}.  
\subsubsection{SCADA Event-Based Process-Control} 
When process events occur, \scada executes event-handling process-control routines. This involves reading and writing relevant parameters to resolve the event (e.g., fault correction, process transition). As such, \scada events can be used to artificially induce its process-control execution, allowing it to be analyzed~\cite{scaphy}.  
\scada events are specified using OPC event tags~\cite{ae, ae1}.
OPC is a widely deployed service in ICS plants to enable data compatibility in common formats called tags~\cite{opc,opc2,opc3}. 
As an example, \fref{fig:opc-setup} shows relevant \scada settings in a real-world chemical dosing process~\cite{scaphy}. \fref{fig:opc-devices} shows the actuators and parameters. \fref{fig:opc-ae} shows the OPC event tags, one of which is \emph{Valve.0} (\intakev). The shown tag, \emph{Valve0.open}, means that \scada responds to device states that trigger the \intakev to open (called \emph{event-trigger} states). 
\fref{fig:stl} shows a function block diagram (FBD) of the process. FBD is a graphical logic that describes how devices are connected and operated. FBD can be converted to readable text, called \emph{statement list} (STL). As shown in \fref{fig:stl}, the STL logic evaluates to \emph{Valve.0}. This means that \emph{Valve.0} is dependent on all devices in the STL block. Hence, the union of all \emph{states} of these devices is the event states \scada responds to. 
\fref{fig:scadabr} shows \scada event-handling settings for the water treatment example.

\subsubsection{Real-World Attack Running Example}
In 2021, an attacker compromised the \scada HMI in the City of Oldsmar's water treatment plant via Teamviewer~\cite{florida,florida2,oldsmartech,attacksok}. He then ran commands to raise the \emph{dosing rate} of Sodium Hydroxide (NaOH) to toxic levels, endangering citizens. Water plants use NaOH to set water PH, but it is toxic in high doses. This \scada attack disrupted the \emph{chemical dosing} process, increasing NaOH from 100 ppm level to 11,100 ppm~\cite{florida,florida2,oldsmartech,attacksok}.
\subsection{Industrial Process Attacks and Challenges}
Modern attacks on industrial processes are launched from infected \scada hosts to alter its parameters~\cite{attacksok2, attacksok3, attacksok4, risi-database, icsstrive}. \scada adversaries evade detection by blending with normal host/network behavior while stealthily perturbing processes, which over time leads to damages~\cite{significance}. Stealthy attacks blend well with benign noise/faults, which triggers false alarms in current process anomaly detection tools. To reduce false alarms, it is vital for operators to connect process anomalies to the causal malicious \scada operation. This requires identifying \scada actions \emph{that have} adverse process effects, and then tying these actions \emph{in time} to the process behavior. This increases detection accuracy and improves attack understanding. Sadly, this is challenging because \scada and industrial processes differ in both semantics and time, i.e., their behavior does not map to each other nor evolve at similar time scales. This is because while \scada executes \emph{discrete} software at CPU speed, processes follow \emph{continuous} physics laws and are subject to \emph{inertial} forces or delays. 

\scada hosts have dedicated connections to devices via secured serial or TCP sessions~\cite{scaphy}. As such, \scada attackers \emph{hitchhike} on their normal operation to avoid triggering host anti-viruses such as terminating the \scada program and initiating a new session with devices. For example, Stuxnet~\cite{stuxnet} infected Siemens \scada to re-program PLCs under the legitimate program context, and the Oldsmar attack used HMIs. Further, due to the presence of re-programmable third-party tools in \scada hosts, defenses such as code-signing may not be easily enforced, allowing attackers to infect \scada programs. 
Unlike in IT, \scada has many complex physical dependencies. As such, applying current program analysis tools~\cite{forecast,s2e} to assess them may be intractable due to hardware-constrained code paths and physical environment need. 
\subsection{Limitations of Current ICS Defenses}
To accurately detect industrial process attacks with low false positives, it is necessary to tie the \emph{causal} malicious \scada operation to the adverse process \emph{effect}. 
Existing techniques inspect data in hosts, PLCs, and physical domains. However, most analyses are done in isolation (i.e., with no cross-domain correlation) as shown  in \tabref{tbl:taxonomy}.
Physical anomaly detection~\cite{tabor, limiting, state, invariants,bro, runtime-powerflow,process-semantics,process-awareness,runtime-monitoring} including reinforcement learning~\cite{rl1,rl2,rl3,rl4} can detect anomalous process behavior but cannot tie it to their \emph{cyber cause}, which leads to false alarms due to benign anomalies such as faults/noise.
PLC defenses \cite{tsv, vetplc} detect altered PLC logic, but are not suited for \scada adversaries, who can modify PLC parameters/coils at run-time without altering its logic. Host anomaly detection~\cite{orpheus, lee} can flag abnormal system calls, but cannot know if the flagged calls have adverse physical effects (i.e., discrete system calls does not map to the continuous process behavior).  Network analysis tools~\cite{dnp_attack, justtraffic, modbusmodel, functioncode, stats1, ml, topologychanges, celine} can detect unusual and noisy traffic (e.g., scans and illegal packet fields) but cannot catch semantic-based attacks (e.g., process perturbation) which can use normal tools to emit normal traffic.
\subsubsection{SCADA and Process Correlation}  
\scaphy~\cite{scaphy} is a recent work that combined \scada and process signatures to detect attacks. \scaphy used OPC events to induce and learn system calls unique to \scada phases. Then, it created physical signatures based on inconsistent device states. However, \scaphy's analysis of \scada and process behavior were disjointed from each other or not connected either semantically or time-wise. For example, discrete calls do not map or capture any process dynamics behavior, which is continuous in nature. This led to high false alarms from its isolated alerts. In addition, \scaphy's design of relying on signatures limits it to only known attacks, i.e., it cannot detect unknown attacks. In our experiments, we show that attacks that use normal system calls (e.g., infected \scada tools) will evade \scaphy's signatures.
\subsection{Threat Model and Assumptions}
We assume a similar threat model as current works where an attacker
has infected \scada programs and can execute commands to alter devices~\cite{invariants, orpheus,scaphy}. Similar to recent work~\cite{scaphy}, \sys comprise of two privileged agents that monitor executions in \scada, and another that logs sensor/actuator values from PLCs (depicted in \fref{fig:connect}/\fref{fig:architecture}). As in these works, the agents are assumed to run in a Trusted Computing Base, hence cannot be tampered with. We do not consider non \scada-originated attacks such as side-channels~\cite{sc-emf,sc-power} or direct physical access to devices/PLCs. Based on public data~\cite{attacksok2, attacksok3, attacksok4, risi-database, icsstrive}, vast majority of in-the-wild attacks are launched from \scada~\cite{scaphy}. We note that PLC Man-in-the-middle attacks have been addressed by previous work~\cite{mitm2, mitm3} and in practice using non-PLC diode gateways~\cite{diode}. Hence it is outside of our scope. 
\subsection{Domain Knowledge Analysis of SCADA and Process Semantics and Time Evolution} 
Our idea to bridge the unique semantic and time evolution of \scada and industrial processes is to first understand process constraints in \scada, and alternatively, identify \scada effects on process dynamics. Then, we use knowledge of how physics factors constrain process time evolution to align it with \scada so their behaviors can be correlated in time. This approach is generic to industrial processes, i.e., only requires logs from actuators/sensors and \scada hosts. Through this, \sys overcomes current limitations in detecting process attacks such as disjointed analysis (i.e., which causes false alerts) and signature-based detection (e.g. as in \scaphy).
\subsubsection{Analyzing Process Constraints in SCADA} Since \scada process-control \emph{adheres} to physics dependencies (i.e., how actuators operate and inter-depend on each other to achieve a physics-governed task) in managing industrial processes, we first used OPC events to induce process-control execution trace, which reveals program-flow dependency of \scada operations on processes. 
For example (without loss of generality), in a chemical dosing process, the \supplyv depends on the \intakev state to operate and transition the water treatment operation~\cite{scaphy}. In the process-control trace, we uncover this by tracing \rread calls on the \intakev followed by a predicate check, and then a \wwrite call on the \supplyv (called \emph{Read-before-Write} process-control constraint). 
Then, \sys measures its continuous statistical behavior such as its frequency and bursts (or intermittence). \sys then aggregates these constraints and uses them  to selectively monitor  the attacker's process-targeted executions that violate intrinsic process constraints, which are as follows: (1) \emph{control-time} or time-interval between two \emph{dependent} \wwrite calls (2) \emph{control-burst} or difference in the number of adjacent \wwrite calls on a single device, and (3) \emph{control-frequency} of device \wwrite calls relative to others. Note that command calls are addressed to devices. Hence dependent commands imply dependent devices. In Windows OS, which dominates SCADA, \wwrite and \rread operations specify the communication session (e.g., serial or TCP) and the Device-Tag in their arguments.

\subsubsection{Embedding Domain Knowledge on Process Dynamics} To analyze anomalous process behavior, \sys develops a physics-informed architecture that embeds knowledge of inertial laws in actuating systems. It uses a PINN that integrates a PDE loss function to minimize inertial irregularities/noise in the process input distribution.  
This increases the learning algorithm's correctness even in limited training data, facilitating it to detect injected subtle adverse effects. Through this, \sys detects malicious \scada effects on processes, especially \emph{stealth} perturbations, which would have been otherwise blended with noisy behaviors in process dynamics. Inertial measurement units (IMUs) in industrial plants can measure inertia delays in actuating systems~\cite{inertia, inertia2}. However, this can also be derived by analyzing the rate of change of process value logs using a data-driven approach.  Specifically, for processes in motion (e.g., a moving conveyor/actuating valve), we measure the average time it takes to come to stop/de-energize after a \emph{stop} command. For processes at rest, we measure the time it takes to attain a steady state (e.g., constant speed of conveyor) after a \emph{start} command
\subsubsection{Capturing Industrial Process Periodicity} 
Industrial actuating systems exhibit periodic/repetitive behavior, such as the up/down movement of a piston, cyclical motion of spinning discs, or off/on switching of regulator valves/pumps. 
Unlike other neural network applications where longer sequences are preferred, intuitively, sequence sizes that capture a process's periodic nature will better learn its behavior and detect anomalous effects. Since inertia captures a process's \emph{reaction time} to changes, \sys uses it to inform its PINN sequence sizes, which aptly fits its periodicity. 
For example, when stop commands are issued to spinning discs, inertial (centrifugal) forces may enable it to complete its \emph{rotation}. Note that this differs from process to process e.g., a short piston may require a shorter sequence to encode its (periodic) up/down motion. To overcome delays in analyzing periodic sequences (i.e., one after another) and prioritize their causal relationships, \sys implements its PINN in state-of-the-art \emph{Transformer} architectures~\cite{transformer}, which learns sequences at once and instantiates multiple \emph{Attention} processes~\cite{attention} for each input in the periodic sequence. 

\begin{figure}[t]
\centering
\includegraphics[width=0.32\textwidth]{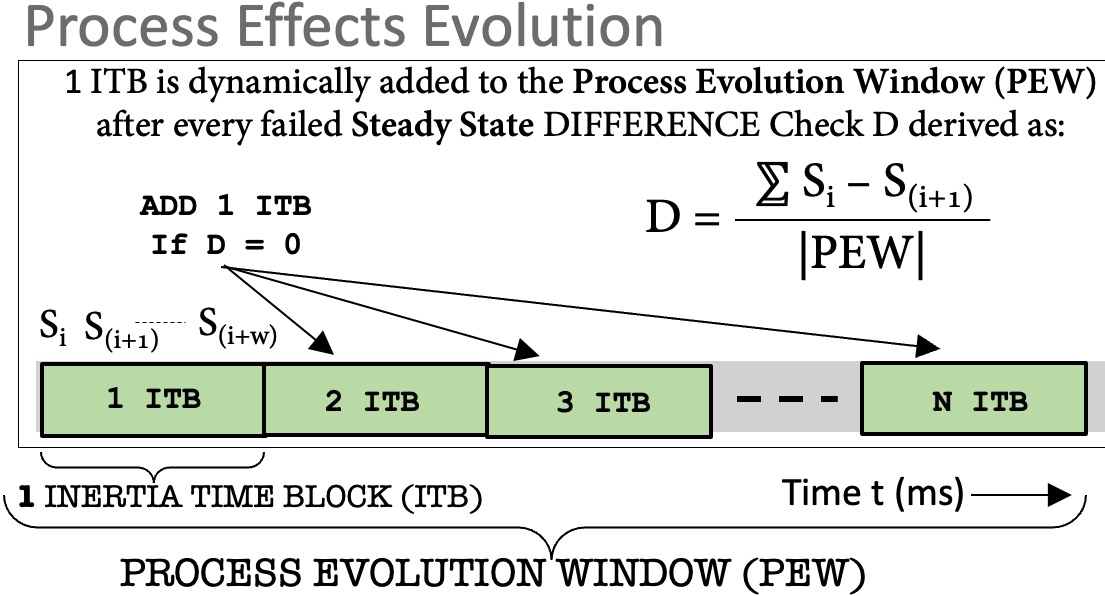}
  \caption{Dynamic Alignment of \scada and Process Time Evolution}
\label{fig:evolution}
\end{figure}
\subsubsection{Aligning SCADA and Process Time Evolution}
Because inertia causes processes to react slower to changes even after an attack, \sys analyzes process inertia delays to inform its correlation of \scada and process time evolution. When a \scada attack is detected, we can expect the process effect to be noticeable \emph{after} the inertia time has passed (called \emph{inertia time block} or ITB in milliseconds). A process behavior within an ITB is captured by  sensor/actuator time series within the ITB period.
After 1 ITB has passed, the lifespan of the anomalous effect (called the \emph{process evolution window}) will vary based on other state conditions. However, it is linear (i.e., non-increasing) for inertial systems~\cite{inertia, inertia2, inertia3}. Hence, we analyze the rate of change of process values to know when it has reached a steady state (i.e. when the avg. difference in successive changes becomes negligible~\cite{scaphy}). We use this to determine when a flagged \scada attack is done having an effect on a process. Since the ITB captures a process' \emph{reaction time} to changes, \sys performs a steady-state checking after each ITB. If the steady state is not reached, \sys adds 1 ITB to the process evolution window and continues to do so until the steady state is reached. This approach is depicted in \fref{fig:evolution}. Through this, \sys dynamically aligns a \scada attack to its resultant process effect, enabling them to be correlated in time.
\begin{figure*}[t]
\centering
\includegraphics[width=0.75\textwidth]{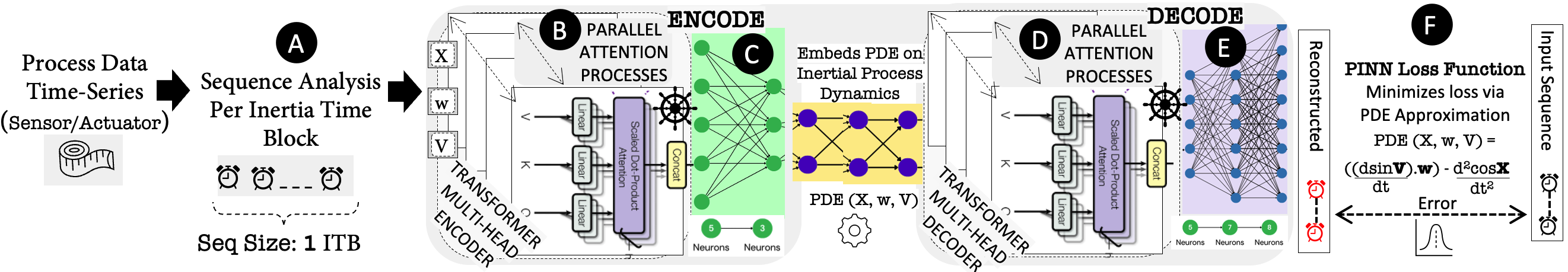}
  \caption{\sys Process-side: Develops a transformer-based architecture that instantiates parallel attention processes for each input sequence. \sys PINN embeds a PDE and loss function to capture inertial properties in process dynamics, facilitating its robustness}
\label{fig:system_physics}
\end{figure*}
 \begin{figure}[t]
\centering
\includegraphics[width=0.45\textwidth]{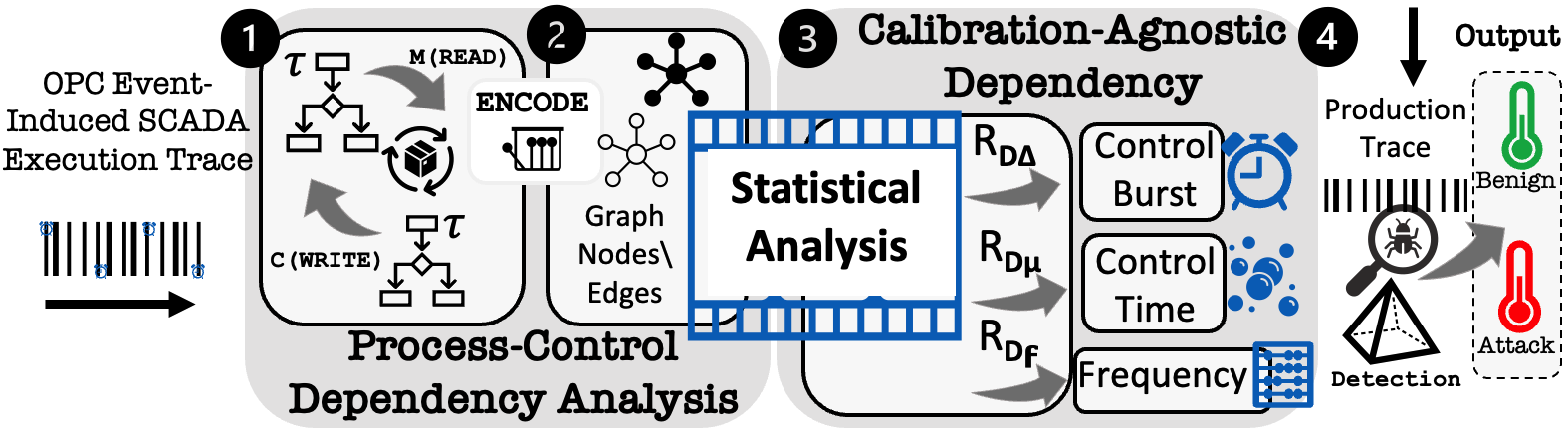}
  \caption{\sys \scada-side: Statistical behavior of process-control dependencies are analyzed and aggregated to derive a calibration-agnostic \rd constraints to detect attacker's violation of contraints}
\label{fig:system}
\end{figure}
\subsubsection{Calibration-Agnostic Constraints}
When a calibrated amount of physical task increases or decreases, dependent devices also increase or decrease their throughput \emph{by the same factor}  to maintain operational dependence~\cite{learn-physics1, learn-physics2}. When this happens, the \emph{absolute} dependency value (e.g., the time interval) between two dependent devices may change, but the \emph{variation} across different task calibrations remains the same.  For example, in a chemical dosing process, no matter how high or low to fill a tank, the ratio of the time interval between \intakev and \supplyv actions vary by the same factor. Therefore, it is possible to derive a single dependency variability that works for any calibration of the same process, as long as the inter-dependent devices remain the same. To do this, \sys leverages an established mathematical foundation called the coefficient of variation (\cv)~\cite{cv, cv2}, which succinctly embodies this physical phenomenon. \cv measures the ratio of standard deviation to the mean to compare \emph{degrees of variation from one data series to another}, even if their means are drastically different. Using \cv, \sys aggregate statistical values for each dependent device pair by computing their \emph{relative dependency} (or \rd) for different runs of the process. Via \rd, \sys process constraints can be applied to detect \scada attacks in a calibration-agnostic way.

\section{Design}
\label{sec:design}
In this section, we first describe \sys's end-to-end approach and its input and output. Then, we describe the formal design of its \scada-side and process-side analysis.
\subsection{End-to-End Approach and Input/Ouput}
At the \scada side, \sys takes as input, OPC event-induced \scada execution trace. It outputs 3 process-control constraints, which map each event to a relative dependency (\rd) measure used to detect \scada attacks, as follows:
 (\emph{I}) event-to-control time (\emph{II}) event-to-control burst, and (\emph{III}) event-to-control frequency maps. An event-to-control time map, $V_{k(i,j)} \mapsto R_{D\Delta(i,j)}$, means that the time-interval $\Delta$ between dependent control commands on devices $i$ and $j$ due to event $k$ in an event set $V$ should not deviate from the benign control time variation $R_{D\Delta(i,j)}$, otherwise, it is anomalous.
 Similarly, event-to-control burst map, $V_{k(i)} \mapsto R_{D\mu(i)}$, means that the control burst between adjacent command bursts on device $i$ due to event $k$ should not deviate from the control-burst variation $R_{D\mu(i)}$.
 An event-to-control frequency map, $V_{k,(i)}\mapsto R_{D\digamma(i)}$, means that the frequency of commands on $i$ due to event $k$ should not exceed the benign control frequency ratio $R_{D\digamma (i)}$. 
\sys \scada analysis works in 4 phases shown in \fref{fig:system}. 
\circled{1} It analyzes the input execution trace and extracts the process-control dependency operations (or \emph{Read-before-Write}). For each dependency pair, \circled{2} it represents their frequency, burst, and time features in a graph, whereby directed edges connect devices in the direction of their dependency relationships. \circled{3} \sys then aggregates these constraints using the calibration-agnostic relative  dependency (or \rd) formulation. \circled{4} During production, \sys analyzes process-control logs in \scada hosts to detect behaviors that violate the \rd constraints.

At the process side, \sys takes as input, time-series of sensor/actuator data per second. It outputs a trained PINN model of the industrial process behavior tailored for correlation with \scada side attacks.
\sys process side analysis works in 6 phases shown in \fref{fig:system_physics}. \circled{A} \sys pre-processes the input sequences (e.g., scaling, one-hot encoding,) and uses the derived inertia delay to inform the sequence size. \circled{B} It then feeds the input into its Transformer-based AE, which uses parallel \emph{multi-head} attention blocks to process each sequence input. The output is then concatenated and \circled{C} compressed via the PINN. At the decode stage, \circled{D} the compressed data is processed again via the multi-head blocks, and then \circled{E} uncompressed. The output is reconstructed and the PINN/PDE loss function is applied to compute the error \circled{F}, and fed back to optimize the training. During production, \sys processes sequences in the same inertia-informed size. As with existing tools~\cite{rnn, rnn2}, it uses the 75th percentile as the threshold to detect anomalies. 
\subsection{Analyzing Process-Control Constraints}
\sys analyzes physical process dependency between two devices $i$ and $j$ based on control commands executed on them, and measures their frequency and time features. It then aggregates their relative statistical behavior (of dependent process-control commands) into 3 categories; control time, control burst, and control frequency.
\subsubsection{Control Command Dependency} Control command dependency ($C_i,C_j$) exists when operation to alter device $i$ depends on device $j$. In \sys, control commands are WRITEs while \emph{Monitoring} are READs. \scada invokes a monitoring command $M_{i}$ to read device $i$ state. 
To formulate control command dependency, we first define a process-control operation $P(V_{k})$ as a combination of control and monitoring commands due to event $k$ in the event set $V$, as follows:
\begingroup
\begin{equation} \label{eq:2}
\footnotesize
P(V_{k}) := \{{ C_{j},C_{j+1},.....C_{n}}\} \cup \{{ M_{j},M_{j+1},.....M_{n}}\}
\end{equation}
\endgroup
A control command $C_{j}$ is dependent on $C_{i}$, denoted as $(C_{i}\hookleftarrow C_{j}$) if device $i$ was the last device to be \emph{read} before device $j$ was \emph{written to}, called \emph{Read-before-Write Adjacency} and denoted with a comma as in ($M_{i}\textbf{,}C_{j}$), and given as:
\begingroup
\begin{equation} \label{eq:4}
\footnotesize
\forall M_{i},C_{j} \in P(V_{k}) \wedge (ts(M_{i}) < ts(C_{j})) : \;\;\;C_{j}\hookleftarrow C_{i}
\end{equation}
\endgroup
$ts$ is a timestamp. This control dependency indicates that any alteration of device $j$ considers the state of $i$ before it is done. \\
Control-time $\Delta$ is the time interval between two dependent commands. \sys computes $\Delta$ as;
\begingroup
\begin{equation} \label{eq:5}
\footnotesize
\forall C_{i},C_{j} \in P(V_{k}) \;\;s.t.\; \;i \ne j:\;\; \Delta(i,j):= ABS(ts(C_{i})-ts(C_{j}))
\end{equation}
\endgroup
where $ABS$ means absolute value. \\
A \emph{burst} $B$ is a continuous execution of commands for the same device, without interruption.
Control-Burst $\mu$ is the difference in \emph{burst size} between two adjacent bursts.
\sys computes $\mu$ for a current burst $B_C$ and previous burst $B_P$ of $C_{i}$, such that ts($B_P$) < ts($B_C$), as follows:
\begingroup
\begin{equation} \label{eq:6}
\footnotesize
(\forall B_{C_{i}}, B_{P_{i}} \in P(V_{k}):\;\;\; \mu_{j} := |B_{C_{i}}| - |B_{P_{i}}|
\end{equation}
\endgroup
Control Frequency $\digamma$ is the number of specific control commands of $C_{i}$ in $P(V_{k})$, computed as:
\begingroup
\begin{equation} \label{eq:7}
\footnotesize
\forall C_{i} \in P(V_{k}) \;\;\; \digamma(i) := |C_{i}|
\end{equation}
\endgroup
\subsection{Calibration-Agnostic Measurements}
We explain how \sys aggregates the statistical behavior of control time, control-burst, and control-frequency using a relative dependency (\rd) modeling to derive process constraints for attack detection regardless of the setpoint calibration of the process task.

\subsubsection{Formulation of Relative Dependency} To formulate \rd, \sys applies a statistical measure of dispersion called the coefficient of variation (\cv)~\cite{cv,cv2}. \cv measures the ratio of standard deviation to the mean to compare
\emph{degrees of variation} from one data series to another, even if their means are drastically different. \sys leverages this \cv property to model dependencies among dependent devices involved in physical task, whereby the \emph{inter-ratio} of work between devices is the same regardless of the calibrated amount of work~\cite{learn-physics1, learn-physics2, limiting}. 
\rd enables \sys to be setpoint-agnostic in detecting  attacks. That is, \sys does not analyze statistical values in absolute terms, (e.g., "time-interval of $70k$ scan cycles is too short"). Rather, it analyzes the inter-dependent variations, or what ensures devices are always working in harmony regardless of the amount of setpoint which can increase/decrease the task duration. Further, devices can have more than one dependency relationship. To adjust for these loose dependency instances, \sys measures 2 properties; the degree of dependency or $\epsilon$, and degree of dominance or $\lambda$ to \emph{adjust} calculated statistical dependency value bounds. Specifically,
$\epsilon$ of a control command dependence $(c_{i}\hookleftarrow c_{j})$ measures \emph{how dependent} a control command $c_{j}$ is to $c_{i}$, based on the ratio of their occurrence $(c_{i}\hookleftarrow c_{j})$ to other $j$'s control dependencies. $\lambda$ of a control burst size of commands $C_i$ measures how \emph{common} or \emph{rare} (i.e., a ratio) a specific control burst size is to other burst sizes.\\
In deriving control-time dependency, let $P(V_{k})$ := \{$C_{i},C_{i+1}...C_{n}$\} be the sequence of process-control commands induced by event $k$ in event set $V$. $\epsilon_{i-1,i}$ is the degree of $(c_{i-1}\hookleftarrow c_{i})$ dependency. $j$ is all instances of time-intervals between subsequent commands for $\Delta_{(i-1,i)}$ of $(c_{i-1}\hookleftarrow c_{i})$ in $P(V_{k})$. $Deviation(j)$ and $Mean(j)$ are the standard deviation and mean of $j$. \sys derives $R_{D\Delta}$ as:
\begingroup
\begin{equation} \label{eq:8}
\footnotesize
R_{D\Delta}(i-1,i) = \frac{Deviation(j) + \epsilon_{(i-1,i)}}{Mean(j)}
\end{equation}
\endgroup
In deriving control-burst dependency,
let $\lambda_{i}$ be the $\lambda$ of $C_{i}$ in $P(V_{k})$. 
\sys derives $R_{D\mu}$ as follows:
\begingroup
\begin{equation} \label{eq:9}
\footnotesize
R_{D\mu}(i) = \frac{Deviation(j) + \lambda_{(i)}}{Mean(j)}
\end{equation}
\endgroup
In deriving control-frequency dependency, 
let |$C_i\in P(V_k)$| be the no. of all instances of command $C_{i}$ in $P(V_{k})$. It derives $R_{D\digamma}$ as:
\begingroup
\begin{equation} \label{eq:10}
\footnotesize
R_{D\digamma}(i) = \frac{|C_i\in P(V_k)|}{|P(V_{k})|}
\end{equation}
\endgroup
\subsection{Anomalous Process-Constraints Violation}
For each event $k$ in $V$,
the \rd model outputs three maps (i) event-to-control time (ii) events-to-control burst, and (iii) event-to-control frequency scores. Event-to-control time map is $V_{k(i,j)} \mapsto R_{D\Delta(i,j)}$,
where $R_{D\Delta(i,j)}$ is the legitimate time-interval semantic dependency in terms of control time of dependent control command pair ($C_{i}, C_{j}$).
Event-to-control burst map is $V_{k(i)} \mapsto R_{D\mu(i)}$,
where $R_{D\mu(i)}$ is the legitimate frequency semantic dependency in terms of control burst of the control command $C_{i}$.
Events-to-control frequency ratios map is given as 
 $V_{k(i)} \mapsto R_{D\digamma(i)}$, where $R_{D\digamma (i)}$ is the frequency semantic dependency in terms of control frequency ratio of $C_{i}$.\\
\PP{Malicious Control Execution}
On observing $C_{i}$, \sys compute the \emph{observed} control time metric $\Delta_{OBSERVED}$:
\begingroup
\begin{equation} \label{eq:14}
\footnotesize
\Delta_{OBSERVED}(C_{i-1}, C_{i}) := \frac{ABS(ts(C_{i-1}) - ts(C_{i}))}{Mean(ts(C_{i-1})-ts(C_{i}))}
\end{equation}
\endgroup
$(C_{i-1} \hookleftarrow C_{i})$. $ts$ indicates timestamp. $C_{i}$ is anomalous if $\Delta_{OBSERVED}$ \emph{deviates} the legitimate control time behavior $R_{D\Delta(C_{i-1,i})}$ mapped to by $V_{k(i,i-1)}$.
\sys will compute the observed control burst metric $\mu_{OBSERVED}$ as follows:
\begingroup
\begin{equation} \label{eq:15}
\footnotesize
\mu_{OBSERVED}(C_{i}) := \frac{|B_{C_{i}}| - |B_{P_{i}}|}{Mean(B_{C_{i}},B_{P_{i}})}
\end{equation}
\endgroup
$C_{i}$ is anomalous if $\mu_{OBSERVED}$ \emph{deviates} from legitimate control burst behavior $R_{D\mu(C_{i})}$ mapped to by $V_{k(i)}$.
\sys computes the observed frequency ratio $\digamma_{OBSERVED}$ as:
\begingroup
\begin{equation} \label{eq:16}
\footnotesize
\digamma_{OBSERVED}(C_{i}) := \frac{|C_{i}|}{|P(V_{k})|}
\end{equation}
\endgroup
$C_{i}$ is anomalous if the frequency ratio $\digamma_{C_{i}}$ \emph{deviates} the legitimate control-frequency behavior $R_{D\digamma(C_{i})}$.


\subsection{Domain Knowledge PINNs For Robust Learning of Industrial Process Behaviors}
PINNs lie at the intersection of neural networks and physical modeling. Network networks loss functions to make robust predictions by learning to find optimal parameters to minimize the value of the loss function~\cite{pinn1}. PINNs use partial derivative equations (PDE) to help the model learn from the intrinsic physical properties of the data at hand. This makes PINNs robust even in partial or noisy data due to the physics-based loss function which provides strong regularization terms that help prevent overfitting~\cite{pinn1,pinn2}.
\sys develops a physics-informed learning architecture tailored to learn and correlate behaviors in industrial processes and \scada. First, it designs a PINN to capture the inertial dynamism of industrial processes via a domain knowledge-based PDE loss function. This limits the space of admissible solutions and facilitates the learning algorithm to converge better and generalize well. \sys then implements the PINN/PDE in a Transformer-based AE to instantiate multiple Attention processes for each process input sequence within an ITB. This enables the PINN to capture the periodicity of process and to timely correlate with the \scada side. 
\subsubsection{Embedding Inertial Properties of Processes}
An AE is a neural network $M$ which consists of an encoder $En(x)$ that compress an input sequence of symbols $x=(x_1, ..., x_n)$ into an encoding $z=(z_1, ..., z_n)$, and a decoder $De(e)$, which reconstructs $x$ from a given $z$ to an output sequence of $y=(y_1, ..., y_n)$. 
When trained with the objective $De(En(x))=x$, an AE learns to summarize the distribution of $x\in X$ where $X\in \mathbb{R}^n$. 
AEs can detect anomalous sequences when trained on a benign distribution because $M$ will fail to reconstruct the features of x~\cite{deepreflect, aesecurity}. To detect an anomaly, given an input reconstruction $M(x)=\tilde{x}$, we calculate the mean-square error (MSE) and check if the output is above a threshold, as follows:
\begingroup
\begin{equation} \label{eq:19}
MSE(x,\tilde{a}) = \frac{1}{m}\sum{}{}(x^i-\tilde{x}^i)^2
\end{equation}
\endgroup
\sys make use of a variational AE, which comprises a generative and latent loss. The generative loss penalizes the model for having differences between the input to the model and the output from the model. We set our generative loss as the MSE loss (or L2-loss). The latent loss compares the learned distribution of input data in the latent layer to that of a Gaussian/Normal distribution. We used the Kullback-Leibler (KL) divergence loss as our latent loss. Then, to embed the inertial properties of processes into our model, we develop and introduce a third term in our loss function based on knowledge of actuating systems under inertia~\cite{inertia, inertia2, inertia3} as follows:
\begin{equation}
\sin\left(\frac{dV}{dt}\right) \cdot \omega - \cos\left(\frac{d^2X}{dt^2}\right) = 0
\end{equation}
where $X$ is a vector of the process' numerical features (e.g., the level of the tank),  $V$ is a vector of the rate of change (with respect to time) of these features, and  $\omega$ is a scaler of their inertia delay. The sin and cosines trigonometric functions capture their angular phase shifts.
The goal of the PDE is to minimize the angular displacement (i.e., bring it to 0) in processes with inertia. To explain, without inertia, the second-order derivative of the angular positions of $X$ (or acceleration) should be equal to the first-order derivative of $V$. That is, $\omega$ is the displacement. 
In order to incorporate the equation into our neural network, we extracted the first term of the equation from the reconstruction of the input data given by the model and the second term directly from the input data. Following this, we set our PINN loss function to be the mean squared difference between the first and second terms. This way, the network can learn parameters that minimize the difference between the two terms. 
\subsubsection{Attention Mechanism for Process Periodicity}
\sys uses the attention mechanism~\cite{attention, transformer} to prioritize important causal relationships in actuators and sensor values in the ITB-sized input time series. In AEs, each step in reconstructing the input is auto-regressive, i.e., previously generated symbols are used as input when generating the next. To mitigate this sequential bottleneck, \sys uses the Transformer multi-head attention architecture~\cite{transformer} to encode and decode each input in parallel. Each encode-decode pass finds a unique attention filter for the same input sequence. 
\sys uses an attention function to compute a \emph{rank} for each input in the sequence by applying the softmax function~\cite{transformer}. For each process, \sys computes the rank on a set of simultaneous queries in the matrix of actuator/sensor vectors, Q, comprised of K and V matrices of their keys and values, as follows:
\begingroup
\begin{equation} \label{eq:18}
rank(Q,K,V) = SoftMax(\frac{QK^T}{\sqrt{d_k}})
\end{equation}
\endgroup
We provide more implementation details in \secref{sec:appendix}

\section{Evaluation}
\label{sec:evaluation}
We evaluate \sys's ability to (i) detect and correlate \scada and process attacks, and (ii) outperform existing works in detection.
\subsubsection{Implementation and Dataset}
\label{ssec:labsetup}
We used public real-world \scada and process data from \scenarios diverse industrial scenarios in water treatment, manufacturing, and shipping plants.\footnote{https://github.com/lordmoses/SCAPHY} Unlike other datasets like SWAT~\cite{swat1} (which only has process data), this data has execution traces of real \scada hosts during the process operations. 
We trained \sys's PINN on a NVIDIA GeForce RTX 2080 Ti GPU ($\sim$300 lines of Python code), which outputs a model of an avg. size of 3-4 MB per process in training time of 1-4 mins per epoch. Each of the process data in the dataset comprises about 4-7 days of continuous process runs. \sys's \scada side comprises of $\sim$240 lines of python code. Both the \scada side and process detection ($\sim$ lines of python code) runs in an Intel(R) Xeon(R) CPU E5-1620 v2 @ 3.70GHz (4 cores). Our implementation is provided online\footnote{https://anonymous.4open.science/r/bridge/}.
\begin{table*}[t]
\caption{Showing \sys's analysis results per process based on public dataset of diverse real-world industrial processes.}
    \centering
    \footnotesize
    \resizebox{0.83\textwidth}{!}{
        \begin{tabular}{@{}|l|c|cc|>{\columncolor[gray]{0.9}}c>{\columncolor[gray]{0.9}}l|>{\columncolor[gray]{0.9}}c|>{\columncolor[gray]{0.9}}c>{\columncolor[gray]{0.9}}c|ccc|c|cccc|@{}}
            \toprule
            Industrial&Diverse& \multicolumn{2}{c|}{OPC Events}& \multicolumn{5}{c}{\scada Process-Control Dependency Analysis} & \multicolumn{3}{|c|}{Process Constraints} &&\multicolumn{4}{c|}{Process Behavior}\\
             \cline{5-9} 
               Processes&Domains&\multicolumn{2}{c|}{(Per Process)}&\multicolumn{3}{|c|}{Process Task and Relevant Actuators}&\multicolumn{2}{c}{Proc Graph} & \multicolumn{3}{|c|}{(Statistical Averages)}& & \multicolumn{4}{c|}{(PINN Training using PDE)} \\  
            \cline{3-12}
            \cline{14-17}
            && \rotatebox{90}{Scenario} \rotatebox{90}{FileSize} &\rotatebox{90}{States/} \rotatebox{90}{Verify} &Description &\rotatebox{90}{Task ID}& Key Actuator  &\rotatebox{90}{CMDs} &\rotatebox{90}{Nodes/} \rotatebox{90}{Edges}  &$R_{D\Delta}$ & $R_{D\mu}$ & $R_{D\digamma}$& \rotatebox{90}{Analysis} \rotatebox{90}{Time (mins)} &\rotatebox{90}{Inertia/}\rotatebox{90}{Seq Size} &\rotatebox{90}{Seq Size} &\rotatebox{90}{Precision} \rotatebox{90}{/Recall} &\rotatebox{90}{F1} \\
            \cline{1-17} 
            &&11.5K &10/0&level control&2.1 &holding tank &24 &4/6 &0.47 &0.125 &0.105 &7.6  &4.5/5&5&85/70&77\\
            \rowcolor{Gray}
           \multirow{-2}{*}{Chemical Dosing}&\multirow{-2}{*}{Water Treatm} &11.5K &14/2& dosing&2.2 &dose valve &22 &5/9 &0.419 &0.111 &0.344 &7.5   &4.5&5&94/87&90\\
            
            &&29K &20/0&pallet alignment&3.2 &Axes X,Z &26 &6 /4 &0.41 &0.133 &0.088 &12.4  &5.1&5&95/93&94\\
           \multirow{-2}{*}{Auto warehouse}&\multirow{-2}{*}{Manufacture} &29K &36/1& throughput&3.2 &entry conveyor &32 &7/5 &0.42 &0.167 &0.034 &12.9   &5.1&6&77/83&80\\
           
           \rowcolor{Gray}
           &&9.5K &28/0&product haul&4.1 &clamp lid/base &22 &5/9 &0.24 &0.71 &0.048 &7.7   &5.7&6&71/73&72\\
            \rowcolor{Gray}
           \multirow{-2}{*}{Assembler}&\multirow{-2}{*}{Manufacture} &9.5K &11/1& load balancing&4.2 &conveyor2 &17 &4/3 &0.686 &0.14 &0.05 &9.3   &5.1&6&88/93&90\\
                        
            &&11K &13/0 &prod safety&11.1 &conveyor1-3 &25 &6/6 &0.198 &0.195 &0.51 &12.0 &7.4&7&92/85&88\\
           \multirow{-2}{*}{Elevator}&\multirow{-2}{*}{Manufacture}&11K &12/0 & throughput&11.2  &entry conveyor &14 &8/5  &0.318 &0.119 &0.691 &6.1   &7.4&7&89/77&83\\

           \rowcolor{Gray}
           &&9K &12/0 &spacing &7.1&buffer conveyor &17 &4/5 &0.091 &0.089 &0.441 &4.7  &6.4&6&82/90&86\\
            \rowcolor{Gray}
           \multirow{-2}{*}{Queue processor}&\multirow{-2}{*}{Manufacture} &9K &29/0& throughput &7.2 &entry conveyor &32 &8/4 &0.358 &0.678 &0.224 &12.1   &6.4&6&88/95&91\\

            &&6K &18/0 &heat setpoint &6.1&room temp &9 &4/5 &0.219 &0.053 &0.09 &11.6  &11.8&13&79/92&85\\
           \multirow{-2}{*}{HVAC}&\multirow{-2}{*}{A/C} &6K &17/1& heat flow&6.2 &vent &23 &5/5 &0.178 &0.410 &0.065 &10.2   &11.8&13&88/83&85\\
            \rowcolor{Gray}
           
            &&7.8K &21/1&load alignment&5.1 &push clamp &33 &9/4 &0.289 &0.845 &0.717 &14.6  &4.1&4&80/88&84\\
           \multirow{-2}{*}{Palletizer}&\multirow{-2}{*}{Shipping} &7.8K &29/0 & prod protection&5.2 &entry conveyor &36 &6/7 &0.091 &0.459 &0.455 &5.8  &4.1&8&71/79&75\\
           \rowcolor{Gray}
            
            \rowcolor{Gray}
            &&6.2K &11/1&path throughput&7.1 &load/unload &32 &7/5  &0.421 &0.616 &0.321 &9.5   &5.1&6&86/93&89\\
           \multirow{-2}{*}{Converge station} &\multirow{-2}{*}{Shipping} &6.2K &19/0& alt throughput&7.2 &transfer &23 &9/8  &0.51 &0.18 &0.085 &11.4  &5.1&6&88/93&90\\
           \rowcolor{Gray}
           &&8.5K &21/1&alignment&4.1 &control arm &22 &8/6 &0.267 &0.371 &0.571 &11.3   &12.1&12&84/99&91\\
            \rowcolor{Gray}
           \multirow{-2}{*}{Production line}&\multirow{-2}{*}{Manufacture}&8.5K &11/0 & throughput&8.2 &conveyors &25 &5/5 &0.211  &0.098 &0.0510 &9.9   &12.1&12&66/94&78\\
            
            &&9K &16/0&sort accuracy&9.1 &unloader &22 &8/7 &0.4 &0.335 &0.118 &15.1  &5.5&6&92/93&92\\
           \multirow{-2}{*}{Sort station}&\multirow{-2}{*}{Shipping} &9K &16/0& throughput&9.2 &conveyor &17 &9/5 &0.107 &0.291 &0.0821 &10.6  &5.5&6&97/70&89\\
           
           \rowcolor{Gray}
           &&4.9K &18/0&accuracy &10.1 &pusher1-2 &26 &11/8 &0.132 &0.439 &0.264 &12.0   &7.6&8&89/73&81\\
            \rowcolor{Gray}
           \multirow{-2}{*}{Separator}&\multirow{-2}{*}{Shipping} &4.9K &15/0& throughput&10.2 &conveyors &27 &8  /7 &0.414 &0.193 &0.072 &9.6  &7.6&8&86/86&80\\
           \bottomrule
        \end{tabular}
    }
    \label{tbl:scenarios}
\end{table*}
\subsection{Modeling and Analysis Results}
\begin{figure}[t]
\centering
\includegraphics[width=0.38\textwidth]{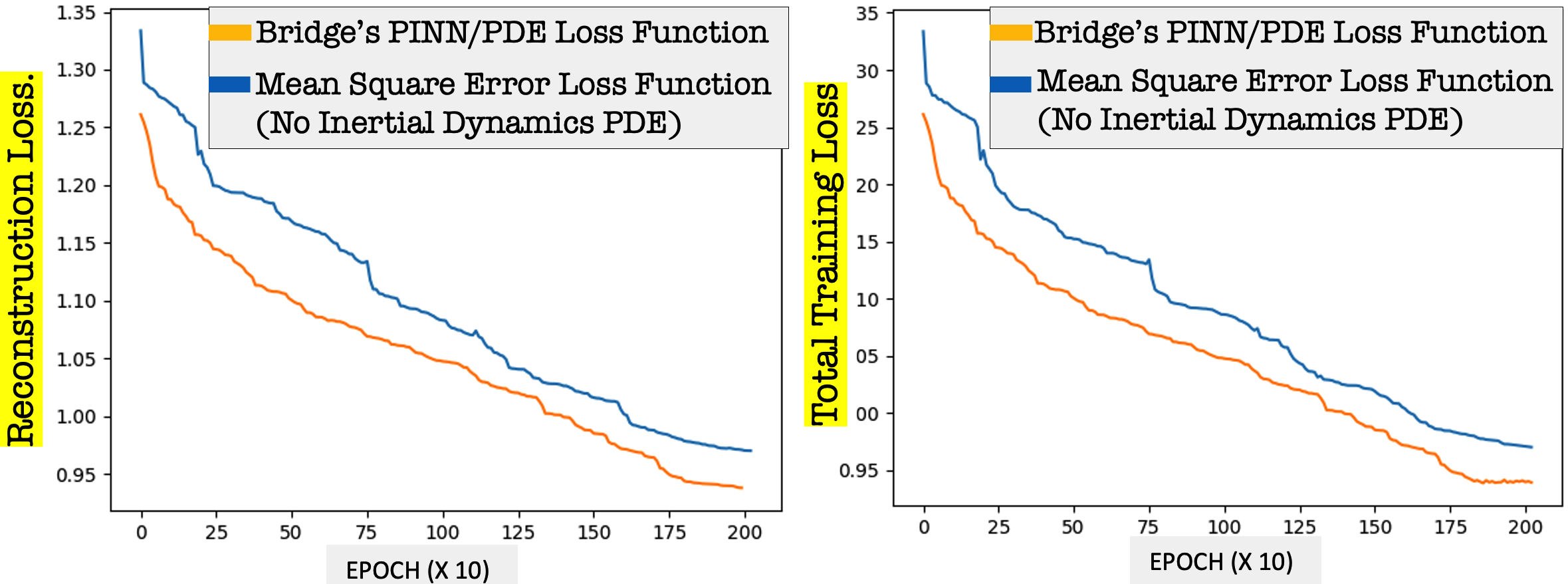}
  \caption{Performance of \sys's PINN vs. current loss functions for learning process behavior: As shown, our PDE loss functions based on Inertial Process Dynamics performed significantly better}
\label{fig:pinn}
\end{figure}

\tabref{tbl:scenarios} presents results from \sys's analysis, at both the \scada and process aspects. We highlight important details. Columns 5, 6, and 7 show the key actuators for each process task.
In column 4 totals for "States", we see that \sys analyzed a total of 338 distinct process-control traces out of the 346 that were induced by the OPC event states (97.7\% relevant behavior coverage). It missed 8 (column 4 totals for "Verify"). This is due to parsing errors when matching PLC/process device names to \scada OPC tags. We verified this via manual checking of the dataset.
In the totals of columns 8 and 9 (i.e., CMD, Nodes/Edges), we see that the 338 traces generated 526 process constraints, 128 of which are from unique device pairs (i.e., total no. of Edges). This shows \sys's robust ability to derive process constraints from process-control executions. Columns 10, 11, and 12 are the statistical relative \rd aggregates of the control-time, control-frequency, and control-burst constraint values. The \scada side analysis took \~8 mins for each process, which is reasonable, based on the scenario sizes (column 3).
The last 4 columns show the PINN training results. \sys attained avg. of training precision and recall above the 85\% mark. Further, the loss curves in \fref{fig:pinn} shows that \sys's PINN performs significantly better than MSE loss functions. The total training loss curve for different ITB runs shows that our model not only has lower training loss but also converges faster than a generic AE. This shows that \sys's PINN is robust. This is also the case in the reconstruction loss (left figure).
\subsection{Adaptive Attacks and Detection Results}
From the public attack dataset~\cite{scaphy}, we draw \attacks attacks from five categories of past incidents as shown in \tabref{tbl:attacks}, which also explains their attack TTP. Each category has 10 attack instances as follows: (\emph{I}) Stuxnet-category ($T831$), (\emph{II}) Industroyer-category ($T855$), (\emph{III}) Oldsmar-category ($T836$), (\emph{IV}) Triton-category ($T801$), (\emph{V}) System crash-category ($T816$). Cat. I-IV attacks target specific actuators or parameters across all processes. They also used a stealthy approach~\cite{incrementattack, limiting, timedelay, timedelay2} to perturb processes. Specifically, for numeric parameters, they steadily increment the process value by a fraction of the current value at each step. For Boolean parameters, the attacks repeatedly toggled them at a slow pace. Cat V attacks (System crash-category) exploit device-specific targets~\cite{icssploit}.  
\begin{table*}
\caption{Attacks Detection Results. Attacks TTPs were generated by recent work based on MITRE/ICSSPLOIT Frameworks~\cite{mitre, metasploit,icssploit}}
    \centering
    \footnotesize
  \resizebox{0.99\textwidth}{!}{
        \begin{tabular}{@{}|l|c|c|l|c|ccc|c|>{\columncolor[gray]{0.9}}c|cc|c|@{}}
            \toprule
            Attack&Total&  && Attacked Process (IDs in \tabref{tbl:scenarios})&\multicolumn{3}{c|}{Process-Control}&Process&&\multicolumn{2}{c|}{}& \\
            Category&Attacks&&Attack  &(2 attack injected per process ID)&\multicolumn{3}{c|}{Violations (Totals)}&Anomalies&Correlated&&&Detect\\
            \cline{7-9}
            &&TTP ID & Description/ TTP Goal& &C-TIME &C-BURST&C-FREQ&(Totals)&Attacks&TP &FP&Time (sec)\\
            \cline{1-13}
           I. Stuxnet Cat.&10&T831 &CTRL Manipulation/Impact Control&2.1, 10.2, 6.2, 9.1, 3.2     &3        & 7    &8  &21  &12&12&0&10.2\\
           \rowcolor{Gray}
           
           II. Industroyer&10&T855 &Unauthorised CMD/Crash Process&1.1, 8.2, 11.1, 7.1, 2.2   && 8    &5     & 19     &10&9&1&7.8\\
           \rowcolor{Gray}
           
            III. Oldsmar Cat.&10&T836 &Modify Parameter/Impact Contr&4.1, 7.2, 10.1, 8.2, 11.2        &17  &11   &   &35  &24&24&0&5.4\\
            
            IV. Triton Cat.&10& T801&  Corrupt State&7.1, 6.2, 8.1, 5.1, 4.2  & 8     &2    & 13       &27      & 22       &22&0& 9.4  \\
             \cline{1-13}    
            &&T816 &Device shutdown/Inhibit Resp fxn&7.1, 9.1, 2, 7.2   &   &1  &   &4    &1&1&0&5.4\\
            &&SPLOIT1.1 &stop Controller/Siemens Simatic-1200 & ICSA-11-186-01(Unprotected Port)&1&1&1&1&1&1&0&7.4\\
            
            V. System-&&SPLOIT1.2 &remote Code execution/QNX SDP 660  &CVE-2006-062(Buffer Overflow)&1  &&1&1&1&1&0&9.2\\
            
            Crash Cat&10&SPLOIT1.3 &remote Device halt/Schneider Quantum &ICSA-13-077-01(I/O corruption)&1&  &1&1&1&1&0&6.7\\
            
            (High&&SPLOIT1.4 &Crash RTOS service/QNX INETDd&CVE-2013-2687(Buffer Overflow)&&1&1&1&1&1&0&7.4\\
            
            Variance)&&SPLOIT1.5 &RPC Device Crash/WindRiver VXWorks& CVE-2015-7599(Integer Overflow)&&1&1&1&1&1&0&12.6\\
            
            Attacks&&SPLOIT1.6 &denial of service/Siemens S7-300/400& CVE-2016-9158(Input Validation)&1&1&1&1&1&1&0&7.2\\
            \bottomrule
        \end{tabular}
        }
    \label{tbl:attacks}
\end{table*}
\subsubsection{Attack Explanation and False Alarms}
For Cat. I-IV, 10 attacks were injected in 5 processes, 2 each. For Cat. V, 4 of the 10 attacks are launched against processes. The remaining 6 use real-world exploit code from ICSSPLOIT~\cite{icssploit}. For example, as shown, exploits of I/O corruption were used do shut down devices. However, because these are \emph{noisy} attacks, they triggered high variance anomalies, especially the control-burst \rdb which is sensitive to chatty behaviors.  In contrast, control-time behavior \rdt detected mostly semantic attacks (Cat I-IV). 
Overall, process anomalies were more rampant than \scada. However, since they are applied as a filter for \scada attack, it did not affect \sys results. 
Although \sys detected all 3 process-constraints violations in \rd, \rdb, and \rdf in many combinations of targets, which also had FPs. However, in some cases, they were discarded since no process effects were seen. Via manual checking, we found that this FP was due to a \scada anomaly  flagged around the same time a process error/noise exist on the system. To mitigate this rare phenomenon, \sys can use the process evolution by discarding process anomalies that started \emph{before} the flagged \scada anomaly. This is because they could not have been caused by \scada (i.e., the effect cannot happen before the cause)
On average, \sys's time-to-detect is about 7-9 seconds (last column), which is practical for near-real time detection.
\subsubsection{Discussions and Lesson Learned}
We discuss how \sys was impacted by the diverse physical processes used. We observe that processes with more actual devices than parameters (such as PID) are less prone to semantic attacks. Control parameters are typically set once during \scada operation, hence having fewer dependencies than actual devices. This is depicted in the no. of generated commands (column 9 of \tabref{tbl:scenarios}). As a real-world example, in the HVAC scenario (which uses soft parameters for A/C control), although 18 states were used to induce \scada, only 9 commands were executed, compared to a more physical system such as Queue Processor, which although generated fewer states, revealed more \scada commands and dependencies. However, processes with more actual devices are more prone to shutdown attacks ($T0816$) than the parameter-based ones. We found that soft parameters allow more automation than mechanical devices. These lessons can inform building secure processes (e.g., resilient to attack).
\subsection{Case Study 1: Real-World Process Attack}
 \label{ssec:florida}
We analyze the 2021 Oldsmar attack, which altered process parameters to dump excess chemicals into the water higher than the calibrated setpoint, while using normal \scada tools/system calls. 
\begin{figure}[t]
\centering
\includegraphics[width=0.34\textwidth]{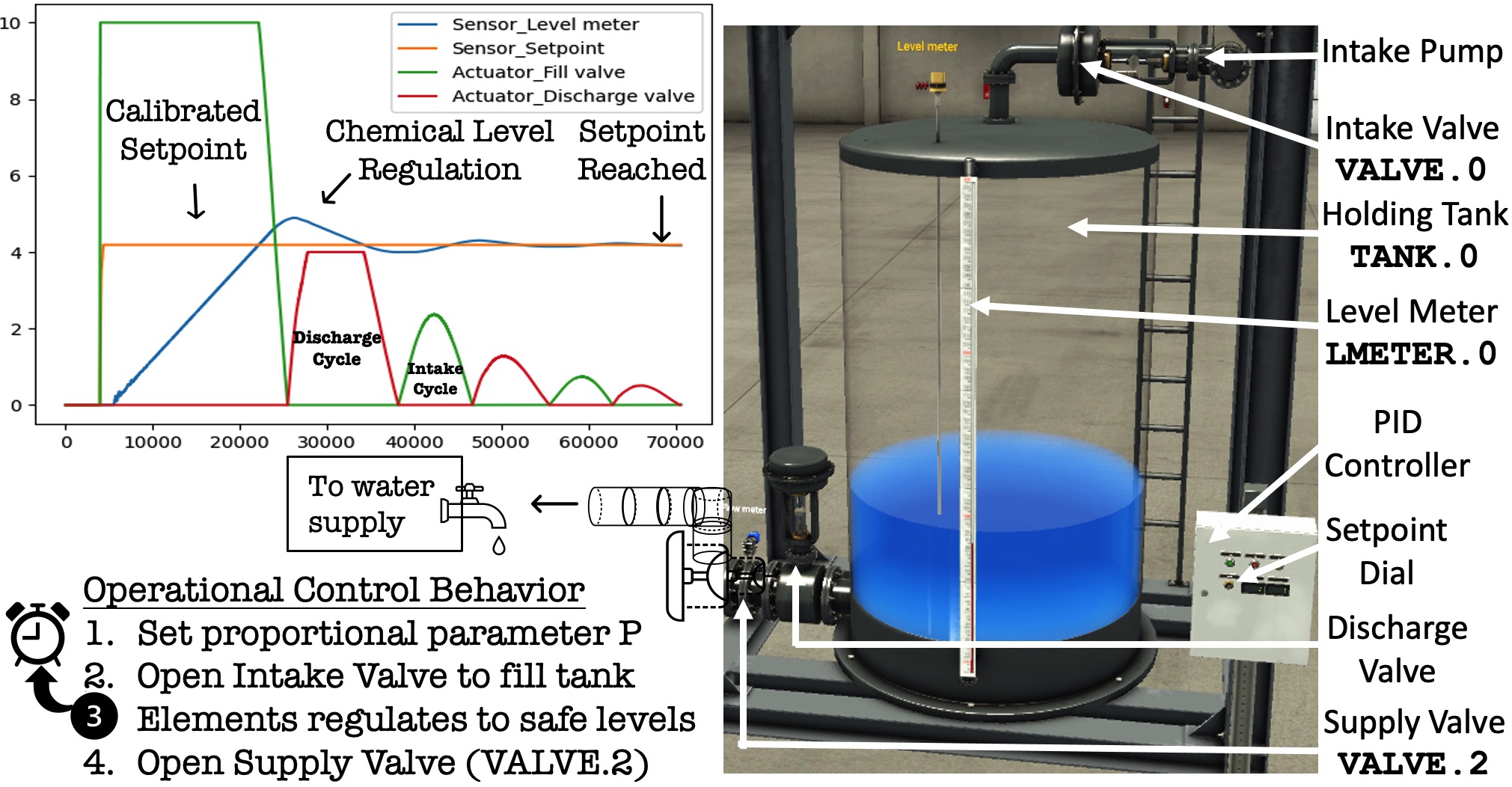}
  \caption{Chemical dosing process constraints violated by Attacker}
\label{fig:motivate}
\end{figure}
\begin{figure}[t]
\centering
\includegraphics[width=0.35\textwidth]{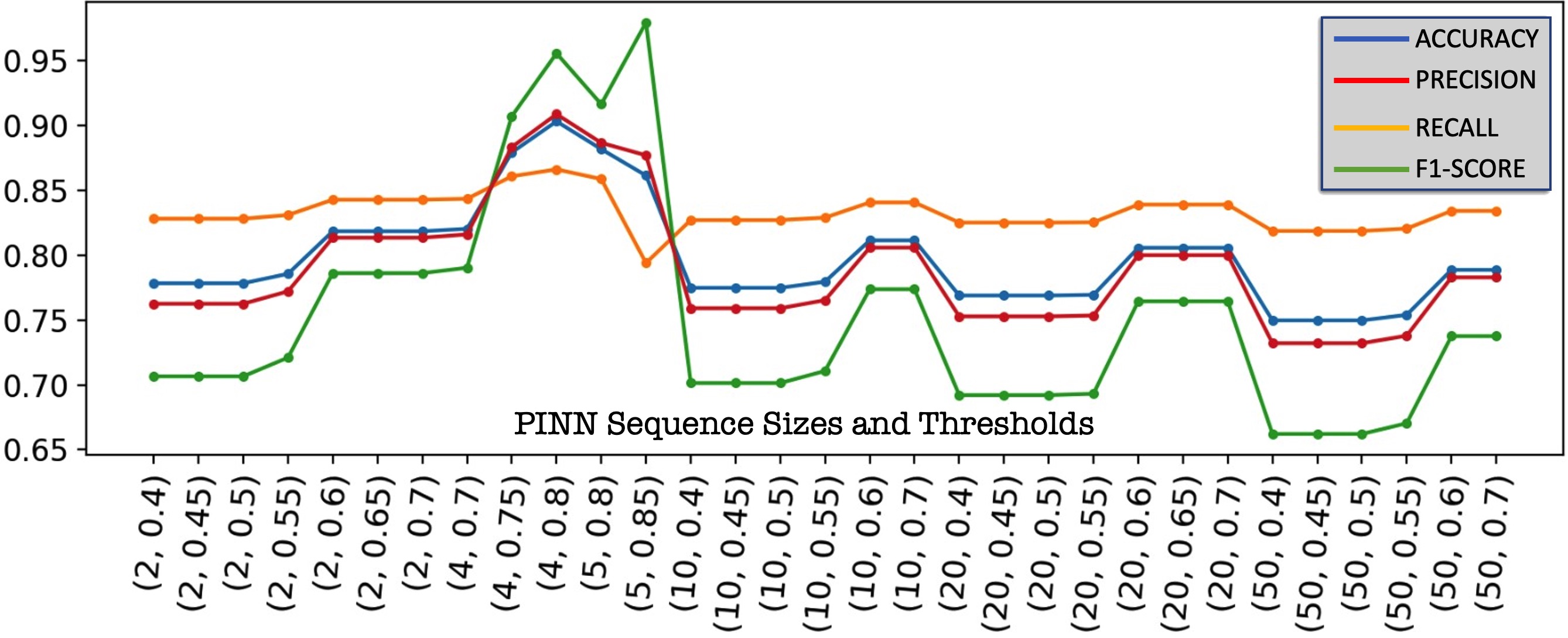}
  \caption{PINN performance for diff sequences sizes. Inertia-informed sizes (4,5 for chemical dosing) performed better overall}
\label{fig:oldsmar-inertia}
\end{figure}
\begin{figure}[t]
\centering
\minipage{0.20\textwidth}%
\begin{subfigure}[c]{\textwidth}
  \includegraphics[width=0.93\textwidth]{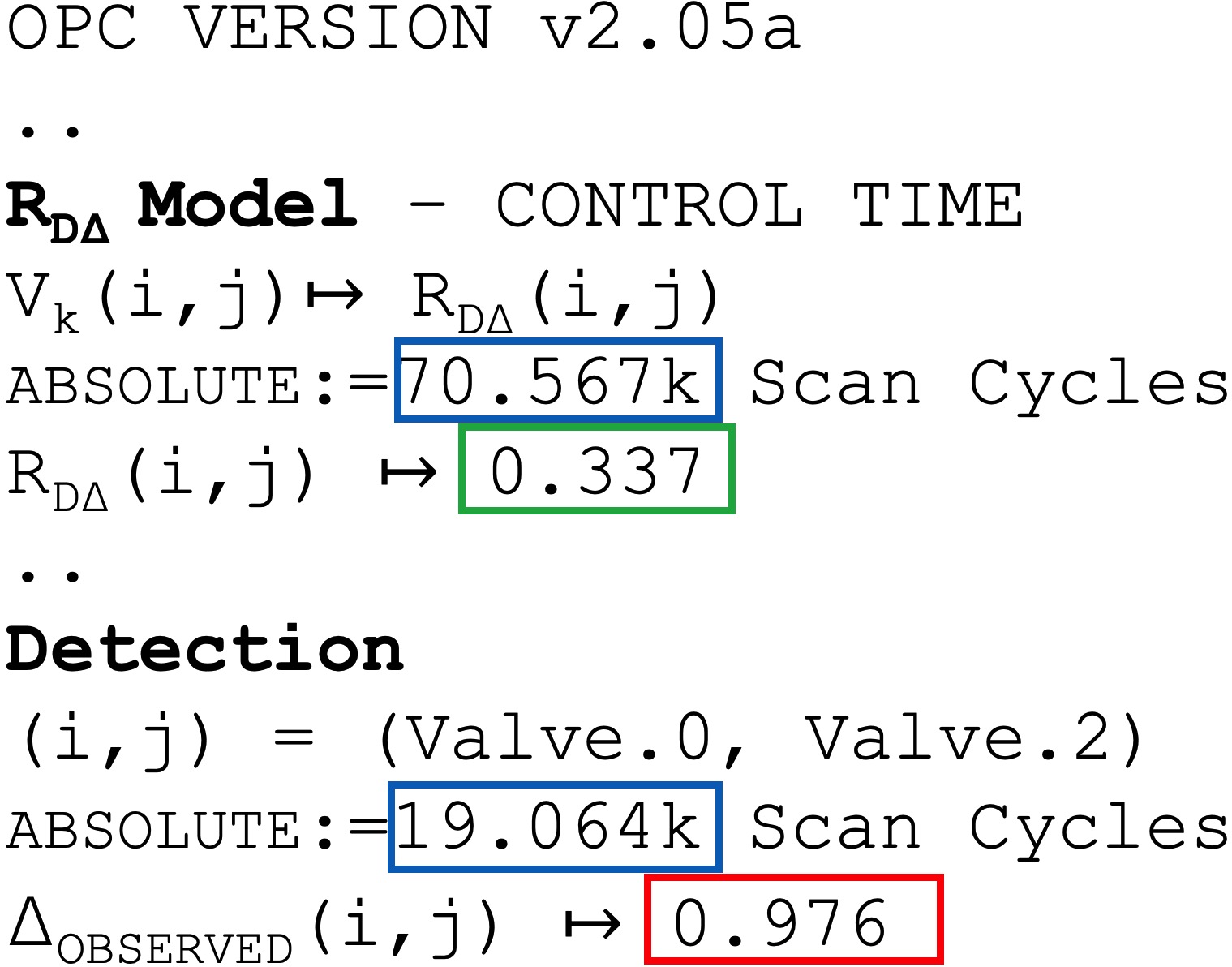}
  \caption{Output Results based on our adapted Oldsmar Example}
\label{fig:oldsmar_output}
\end{subfigure}
\endminipage\hspace{.5em}%
\minipage{0.20\textwidth}%
\begin{subfigure}[c]{\textwidth}
pp  \includegraphics[width=0.98\textwidth]{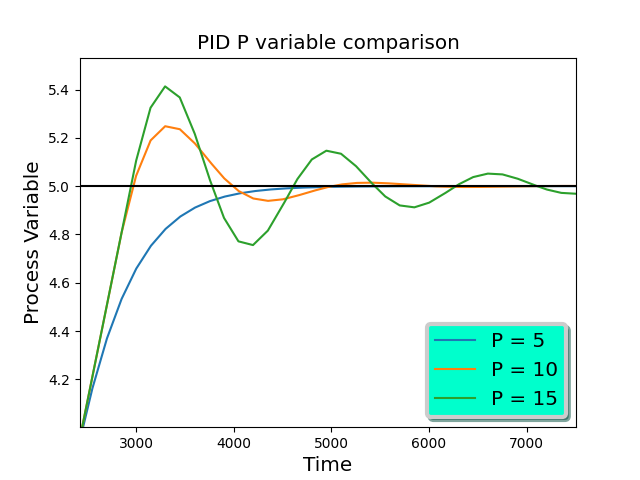}
  \caption{Effect of different $P's$ to reach \sv}
  \label{fig:pchanges}
 \end{subfigure}
\endminipage
\caption{Depicted output for our running example attack, and the effect of different $P's$ to reach the calibrated setpoint}
\end{figure}
\begin{figure}[t]
\centering
\includegraphics[width=0.40\textwidth]{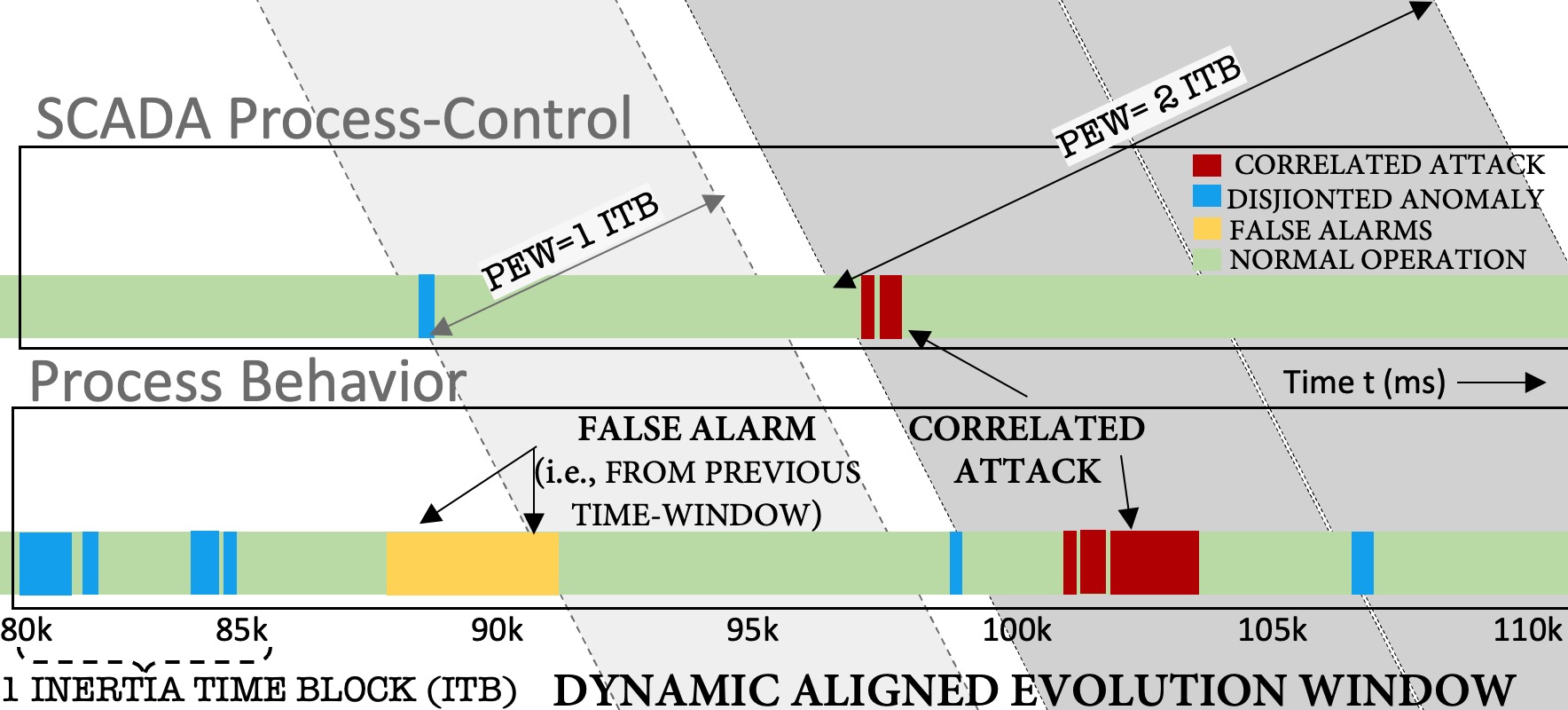}
  \caption{Example anomaly correlation of our running example}
\label{fig:inertia}
\end{figure}\\
\subsubsection{Chemical Dosing Operation and Injected Attack} Chemical dosing involves two processes: \emph{Level control} (LV) and \emph{Dosing}. \fref{fig:motivate} shows the HMI and main devices involved. LV regulates chemical in a holding tank, \emph{Tank.0}, based on \sv. When \sv is reached, \emph{Dosing} opens the \supplyv, \emph{Valve.2} to let the chemical into the water supply. Level control is based on a physical domain \emph{Proportional Integral Derivative} (\emph{PID}) logic. Because \sv cannot be reached in one shot, \emph{PID} performs several "intake" and "discharge" cycles, (\fref{fig:motivate}) whereby an Intake Pump \emph{Pump.0} and \intakev \emph{Valve.0} fills chemical into \emph{Tank.0}, and a Discharge Valve \emph{Valve.1} remove excesses, until \sv is achieved, shown in \fref{fig:motivate}. A PID parameter $P.0$, controls how aggressively the intake and discharge cycles are driven. E.g., a high $P.0$ pumps an \emph{initial} excessive volume into \emph{Tank.0}. \fref{fig:pchanges} shows how different $P.0$ values affect how \sv is reached. \sv is set in the PLC hardware dial, so cannot be altered via \scada. 
To perform the attack, the attacker set $P.0$ to a high value, set \emph{Valve.0} to open, and then, opened \emph{Valve.2}. Although this execution sequence is normal, the control time between opening the \fillv and \dosingv is less than the operational time for LV to regulate the volume. This caused the initially filled excessive volume due to $P$ to be dumped into the water supply.
\subsubsection{Evading Existing Techniques} The \scaphy work noted that it could not detect the \scada attack since benign system calls/HMI were used, which evades its signature approach~\cite{scaphy}. On the process side, \scaphy flagged when the process went over the \sv. However, we see that the attacker can stay just below the \sv, but over time causes the same damages.
\feng misclassified the high $P$  value as an attack, which is a false alarm. Note that although the very high $P$ value seems anomalous, by itself, it is benign because with time, LV will regulate excessive volumes to safe levels. In addition, the attacker could incrementally perturb $P$ to be just below \feng's threshold, while still achieving the attack. In this case, since all other actuator sequences will be normal, \feng will miss the attack.
\begin{table*}[t]
    \centering
    \caption{This is to show \sys's Resiliency to configuration changes using randomly and uniformly generated setpoint \sv for the Chemical Dosing scenario. We show (1) \sys's $R_{D}$ Dispersion  based on $(i,j)=(Valve.0, Valve.2)$ and $(i)=P.0$ and (2) Detection Results}
    \small
    \resizebox{0.75\textwidth}{!}{
        \begin{tabular}{@{}l|cc|ccc|ccccc@{}}
            \toprule
            &Max $S_{V}$            &Min $S_{V}$ &\multicolumn{3}{c|}{$R_{D}$ Average /  $R_{D}$ Dispersion} & \multicolumn{5}{c}{\sys Detection Resilience}\\
            &Config &Config   &$R_{D\Delta(i,j)}$ &$R_{D\mu(i)}$ & $R_{D\digamma(i)}$ &Total  & Detected & Missed & TP\% & FN\% \\
            \cline{1-11} 
           Normal Calibration &2.35&2.35&0.    0.092/-         &0.092/-            &0.25/-    &1 &1 &0 &100 &0 \\
            \rowcolor{Gray}
            Random Calibration Changes &4.6&1.7        &0.187/0.303   &0.118/0.3744       &0.25/0.443 &53 &51 &2 &96.2\% &3.8 \\
            \rowcolor{Gray}
            Uniform Calibration Changes &10.0&1.0      &0.293/0.427     &0.156/0.418  &0.25/0.593      &53 &48 &5 &90.6\% &9.4 \\
           \bottomrule
        \end{tabular}
    }
    \label{tbl:configchanges}
\end{table*}
\begin{table}
\caption{Comparison \sys with Existing Approaches}
    \centering
\resizebox{.97\columnwidth}{!}{
        \begin{tabular}{@{}l|l|c|c|c|ccc@{}}
            \toprule
             Reference&  &Attack/&Semantics&Time&\multicolumn{3}{c}{Detection Results}\\
             \cline{6-8}
             Work & Approach &Benign&Correlated&Correlated&TP    &FP    &FN\\
            \cline{1-8}
             \rowcolor{Gray}
             \scaphy& \scada + &120/&0 (0\%) &71(75\%)        &94 &15 &26   \\
                \rowcolor{Gray}
            ~\cite{scaphy}&Process &268&(disjointed)&(with maps)&\existingaccuracy &\existingfp&21.6\%  \\
            \feng&  Process &120/&-&-     &89    &17     &31   \\
              Rules~\cite{invariants} & Invariant &268&N/A&N/A&74.2&16\%&25.8\% \\
             \rowcolor{Gray}
             &\scada +  &120/&119 (100\%) &119 (100\%)   &118  &1     &2            \\
             \rowcolor{Gray}
             \multirow{-2}{*}{\sys} &Process &268&(false=1)&(Avg. ITB=3)&\accuracy&\fp&1.6\%\\
           \bottomrule
        \end{tabular}
        }
    \label{tbl:comparison}
\end{table}
\subsubsection{\sys Analysis} \sys derives the  chemical dosing inertia via the de-energizing time of \emph{Pump.0} and \emph{Valve.0}, which took about 4.7 sec before the tank stops filling. Based on this, the ITB is 5. \fref{fig:oldsmar-inertia} shows the PINN performance of the with different sizes. As shown, several thresholds on sequence size 5 achieved the best results, which supports our hypothesis. 
The attack was injected around the 97-98th sec as shown in \fref{fig:inertia}. 
It blends with normal operation by listening for the same events as \scada, which is \emph{Valve.0.Open}. 
\subsubsection{Correlation and Detection}
In \scada, \sys detected an anomalous \emph{control-time} dependency between \emph{Valve.0} and \emph{Valve.2}. The output derivations are shown in  \fref{fig:oldsmar_output}. \sys then waited for the inertia time of 4.7 sec before analyzing its process output. Steady-state was reached in one inertia time block (that is, only one inertia time block was added to the process evolution window). \sys flagged an anomaly around the 101th sec. Since both anomalies exist on the correlated \scadaphysics time window, alarm is raised as an actual attack. 
\fref{fig:inertia} shows several isolated anomalies of the process (in blue). We captured a rare occasion around the 90th sec when false alarms arose in both \scada and Physics. However, \sys filters it as a false alarm since the process anomaly permeated from the previous window (88th sec). As such, the process anomaly may not have not come from \scada (i.e., the effect cannot happen before the cause).
The chemical dosing operation was induced by event $V_{k}:=$\emph{Valve.0}.$Open$, via a \rread(tag.Valve.0) API call. \sys detected calls that opened the Intake Valve as \wwrite(tag.Valve.0) (which begins the filling) and the Supply Valve opening as \wwrite(tag.Valve.2). That is, $(i,j)=(Valve.0, Valve.2)$. \sys then computes the \emph{observed} time-interval for the dependent commands on $i,j$ as $\Delta_{OBSERVED}(i,j)$ := $0.976$, which deviates from the established $R_{D\Delta(i,j)}\mapsto 0.337$, hence anomalous. Note that the absolute time-interval of 70k and 19k scan cycles shown in \fref{fig:oldsmar_output} is to illustrate time interval in absolute terms. However, \sys computes relative dependency measure (i.e., \rd= 0.337).
\sys's alert to operators contains the affected devices (e.g., the \supplyv and \intakev tags). This helps them better understand the attack and inform threat remediation at both \scada and physical.
\subsection{Case Study 2: Real-World Mimicry Attack}
\label{ssec:evade}
\begin{figure}[t]
\centering
\includegraphics[width=0.31\textwidth]{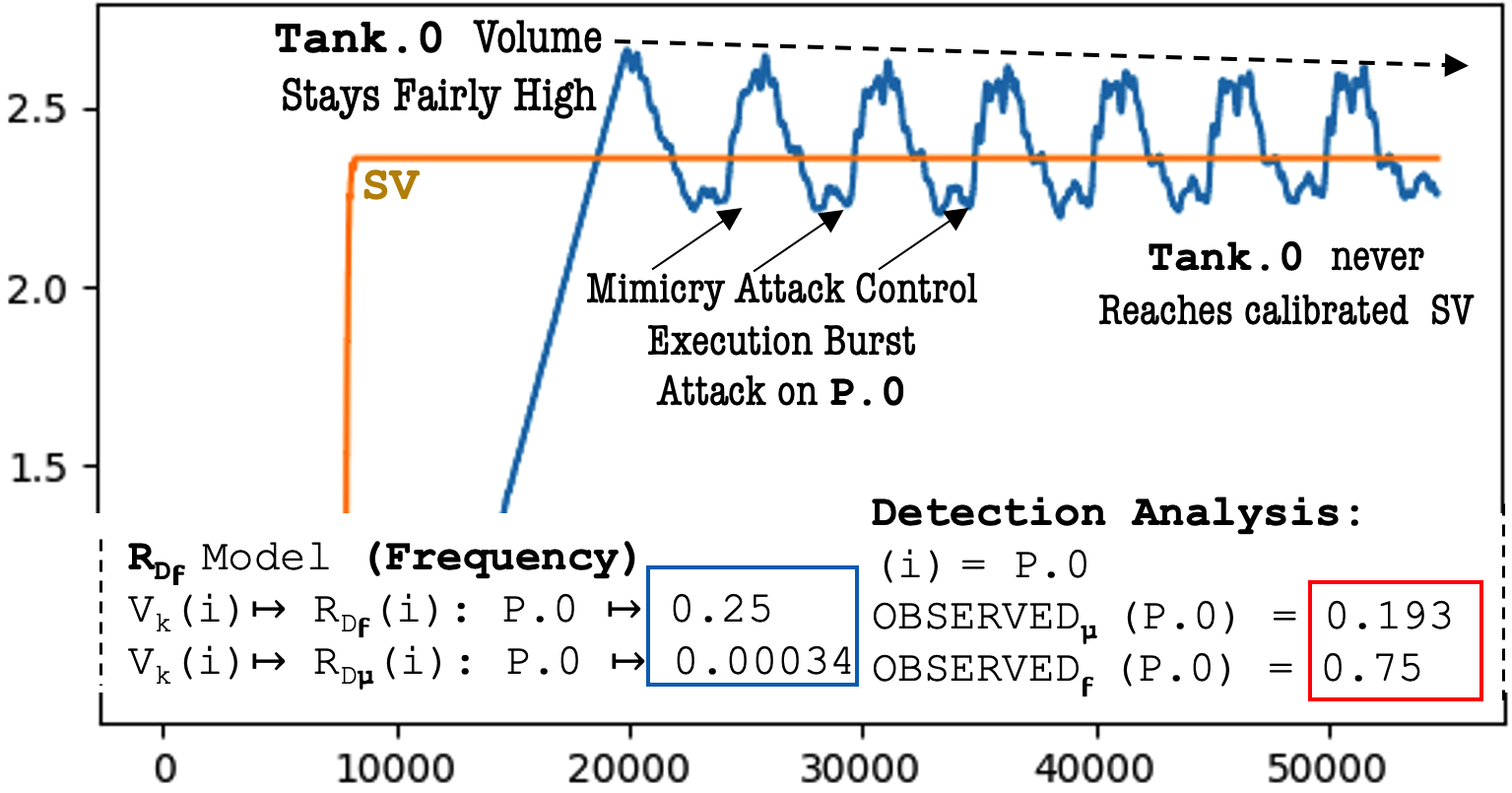}
  \caption{An Evasive Mimicry Oldsmar Attack: Can evaded \rdt, but violates \rdb \rdf behaviors, allowing \sys to detect it}
\label{fig:physics_malware}
\end{figure}

We used an advanced Oldsmar attack that mimics \sys's timing \rd dependency behavior to evade \sys and achieve its attack. This attack obeys the expected control-time between \emph{Valve.0} and \emph{Valve.2} by carefully incrementing parameter $P.0$ every few scan cycles, such that as the process gets close \sv, process-control (based on $P$) still pumps excessive intake chemical. The attack uses the same semantic logic as process-control to
monitor the difference between $L.Meter.0$ to \sv to calculate how much to increment $P$ to keep $Tank.0$ high enough until \sys's control-time is satisfied. To explain this semantic attack, recall that \emph{PID} process-control logic uses $P.0$ to compute how aggressively to fill the tank at every process (i.e., intake and discharge) cycle, which is about 9k scan cycles based on \fref{fig:motivate}. Given that \emph{Level control} takes 70k scan cycles to achieve \sv, there's room to manipulate $P$ to still poison the water. The attack commands kept $Tank.0$ high enough as shown in \fref{fig:physics_malware}, and when \sys's expected \rd is reached, it opened the dosing valve \emph{Valve.2} achieving the same attack goal. However, \sys detected the attack based on control-burst \rdb and control-frequency \rdf behaviors, shown in \fref{fig:physics_malware}. The attack incremented $P.0$ a total of 18 times out of 24 total commands leading to an anomalous \emph{control frequency ratio} of 0.75 compared to $R_{D\digamma}=$0.25. Further, the injection bursts of 2,3,2,3,2,3,3 led to a control burst behavior of 0.193 compared to $R_{D\mu}$=0.00004 (i.e., $C_{P.0}$ almost always have burst of 1). These \scada anomalies were validated by process in the same time window, but may evade isolated sensor analysis.
\subsection{Case Study 3: Process Calibration Changes}
We draw experiment data to evaluate \sys resilience to \sv changes. Recall that \sys's \rd model is designed to work for different calibrated \sv.
In this experiment, we answer 2 questions (i) how much does \sys's \rd \emph{disperse} when derived from varying \sv of the same CPS scenario, and (ii) how many attacks are \emph{missed} due to new \sv. We chose the chemical dosing scenario for this experiment because it provides many \sv decimals, from which we generated 53 \sv calibrations. \tabref{tbl:configchanges} details our results. For (i), columns 4-6 show the average output of $R_{D\Delta}$, $R_{D\digamma}$, and $R_{D\mu}$, which are very close to their original derivation. To mathematically reason about their \emph{dispersion}, we compute their standard deviation (SD). As shown, the low SDs indicate very low dispersion, showing that \sys's \rd is very resilient to \sv changes. For (ii) columns 5-8 show \sys detection performance in new \sv configs, with high 96.2\% and 90.6\% TP for \emph{randomly} and \emph{uniformly} generated \sv changes. We found that slightly higher FN in uniformly generated changes is due to boundary limits rarely seen in the plant. We found that we can tune this behavior using \sys's degree of dependence and dominance ($\epsilon$ and $\lambda$), which are designed to tune dependencies that rarely occur.
\subsection{Comparing Against Existing Work}
\label{ssec:compare}
We compare \sys with current approaches as shown in \tabref{tbl:comparison}, which are \feng~\cite{invariants} and \scaphy~\cite{scaphy}. \feng mined invariant rules from predicates generated in sensor traces. These rules must be satisfied between the predicates and actuator states otherwise the states are deemed anomalous. 
\scaphy detects attacks in both \scada and processes. It learns system call signatures unique to \scada phases and created process signatures from observing disruptive physical impact.
To compare \sys against these works, we used the public dataset in~\cite{scaphy}.
\scaphy's implementation is online. For \feng, we used an open-source tool~\cite{invariant-tool} to first build a network graph of each process' FBD/STL, and then generate rules from the time-series of sensor/actuator states that satisfy them.
\subsubsection{Results}
We draw experiments from the dataset whose behaviors are observable from both \scada and process traces, totaling 120 attacks and 268 benign instances. For the detection threshold, we used normalized average values based on the 75\% percentile. \tabref{tbl:comparison} shows the results. Overall, \sys achieved a \accuracy detection accuracy with \fp FP, compared to \existingaccuracy detection and \existingfp FP of \scaphy, which had the second-best result. \sys correlated all its detected attacks in both semantics and time (columns 4 and 5), with an average process evolution window of 3 ITBs for each process. Although \scaphy had 0\% semantic correlation due to it's inherent design flaw, when we provide it with our ITB evolution mappings, it correlated 75\% of its detected attacks in time (column 5). \scaphy missed many attacks due to it's use of known signatures. Further, its use of "outside setpoint" process anomaly signature caused false alarms. \feng detected 89 attacks, i.e., 74.2\% true positives (TP) with 17 FPs (16\%). The high FP is due to flagged sensor deviations that are part of normal behaviors but misclassified by rules. For example, recall in Oldsmar, a high $P.0$ by itself is a benign behavior, but only malicious during the transition to the dosing task. In addition, rules constructed by invariants can be incomplete based on the training data~\cite{ssa}. Further, since \feng is a process-only technique, it did not perform correlation with the \scada side, which impeded it's detection results.
\subsubsection{Sensor Invariant Rules vs. Process-Control Constraints}
We drew a general experiment to compare \sys's process (dependency) constraints (i.e., how devices depend on each other to achieve a task) 
with invariant rules (i.e., properties that should hold in sensor readings). To draw up this experiment from the public data, we injected random perturbations in a normal process and 
\scada data of the water treatment scenario, similar to what was done in the~\cite{invariants} work. Results are shown in \tabref{tbl:comparison2}. For our evaluation metric, we derived a \emph{Noise-Ratio} score by dividing the FP with the total dependency or rules generated. Overall, constraints from process-control  had better results with low Noise-Ratio (15.1,9.6,12.5) compared to (41.4,30.6,41.3) of invariant rules. Intuitively, invariant rules may have \emph{hard} predicates which semantic attacks can still satisfy while causing disruption~\cite{ssa}. For example,\feng assumes that every update of actuator states is due to critical predicates.~\cite{debunk, ssa} showed that such rules may be spurious or incomplete. In contrast, process-control behavior is \emph{task-driven}, hence deviations from them have a high chance of violating intrinsic constraints.

\subsection{Limitations and Discussions}
\label{sec:discussion}
\subsubsection{Manual Effort Required}
Due to the namespace difference between \scada OPC tags and process PLC/device names, we had to manually inspect the PLC FBD/STL files (i.e., shows how devices are connected) of each process to match process-side device names with the \scada side. However, in some scenarios, the namespace was similar (e.g., "VALVE.1" vs. "VAL-1"), which enabled us to use a semi-automated regular expression matching script. This manual activity took us under an hour to match device names in each process. However, it is only to be performed once per scenario.
\subsubsection{Physical Changes} Since \sys relies on physical dependencies, adding new devices will break its process constraints. Since \sys is mostly automated, new  dependencies can be re-learned.
\subsubsection{Correlation Overhead}
When \scada attacks are flagged, \sys must wait for at least 1 ITB to check for the process effect. Although inertia helps to prevent instant damages, depending on the sensitivity, damages may occur before operators can respond. While this limitation applies to all process anomaly tools, we can explore early warning signs. \sys detects attacks around 8s on avg., which is useful for near real-time detection (\secref{sec:evaluation}).
\section{Related Work}
Unlike in IT, industrial processes are governed by physics laws, whose normal behavior can be learned to detect attacks~\cite{learn-physics1, ar, state, limiting, ssa}. In the wild, cyber-attacks originate from infected \scada hosts, which can inject stealthy process changes, but overtime leads to damages. Process anomaly techniques~\cite{bro, ar, state, limiting, ssa, runtime-powerflow,process-semantics,process-awareness,runtime-monitoring} such as \feng and Tabor~\cite{invariants, tabor} can detect process (sensor) changes but cannot identify the \scada cause, hence have false alerts due to noise/faults~\cite{criticalstate}. Host anomaly tools such as Orpheus and Lee~\cite{lee, orpheus} can detect abnormal system calls, but can't know if it caused adverse process effects, which also causes false alarms. To counter false alarms, \scaphy~\cite{scaphy} combined \scada and process anomaly data, but limited by both signature-based detection and false alarms because \scada execution does not map to process dynamics nor evolve at similar time-scales (i.e., semantic/time gap). \sys uses domain knowledge in \scada  process-control and process-tailored PINNs to bridge their unique semantic and time gap, which reduce false positives and improve attack understanding. 

\begin{table}
\caption{Sensor Invariant Rules vs. Process-Control Constraints}
    
\resizebox{0.47\textwidth}{!}{
        \begin{tabular}{@{}l|>|cc|cc|c|>{\columncolor[gray]{0.9}}c>{\columncolor[gray]{0.9}}c>{\columncolor[gray]{0.9}}c>{\columncolor[gray]{0.9}}c|@{}}
        \toprule
            &\multicolumn{4}{c|}{Physical Actuator States Perturbed}&&\multicolumn{4}{c|}{}\\
            \cline{2-5}
             &\multicolumn{2}{c|}{Fill Valve}&\multicolumn{2}{c|}{Discharge Valve}&&\multicolumn{4}{c|}{Results}\\
             \cline{7-10}
             
             Approach&\rotatebox{90}{level} \rotatebox{90}{(volts)} &\rotatebox{90}{Flow} \rotatebox{90}{(cm/$s^2$)})&\rotatebox{90}{level}\rotatebox{90}{(volts)} & \rotatebox{90}{Flow}\rotatebox{90}{(cm/$s^2$)}&\rotatebox{90}{No. Of} \rotatebox{90}{Trials}&\rotatebox{90}{DEPS/} \rotatebox{90}{RULES}&FN&FP&\rotatebox{90}{Noise} \rotatebox{90}{Ratio})\\ 
             \cline{1-10}
             &@5&0.984&@5&-0.975&100&41&11&17&41.4\%\\
             Invariants&@1&0.1962&@1&-0.184&100&62&8&19&30.6\%\\
             \multirow{-1}{*}{Rules}&@10&1.968&@10&-2.025&100&29&12&12&41.3\%\\
             \cline{1-10}
             &@5&0.984&@5&-0.975&100&33&7&5&15.1\%\\
             Process&@1&0.1962&@1&-0.184&100&52&2&5&9.6\%\\
             \multirow{-1}{*}{Constraints}&@10&1.968&@10&-2.025&100&24&10&3&12.5\%\\
           \bottomrule
        \end{tabular}
        }
    \label{tbl:comparison2}
\end{table}
PLC Man-in-the-Middle (MITM) attacks such as Harvey~\cite{harvey} can present false sensor data to \scada.~\cite{mitm2,mitm3} proposed MITM mitigation such as using separate sensor channels channels and physics knowledge.~\cite{diode} uses non-PLC diode gateways to avoid MITM. VetPLC and TSV~\cite{vetplc, tsv} can detect  PLC logic alteration, but not suited for \scada attack which can alter PLC I/Os at runtime without touching its logic. Traffic analysis~\cite{dnp_attack, celine, telemetry, modbusmodel, justtraffic, stats1, ml, topologychanges, pattern2, modbusmodel,criticalstate, pattern1} are promising for abnormal traffic such as scans, but evaded by modern attacks which are semantic-based~\cite{sok}.
Reinforcement and Deep Learning~\cite{rl1,rl2,rl3,rl4} uses game theory to learn attack behaviors but requires known attacks and expert reward function, which may limit its diverse use in ICS. \sys \emph{generalizes} attacks by detecting anomalies that \emph{violates} intrinsic dependencies in \scada and process operation.

\section{Conclusion}
\label{sec:conclusion}
\sys is a domain knowledge-based correlation of \scada and industrial process behaviors that bridges their unique semantic and time evolution differences in detection ICS process attack. This reduces false alarms and improves attack understanding. \sys achieved high accuracy and outperformed existing work.

\bibliography{bib/refs}
\section{ Appendix}
\label{sec:appendix}

\subsection{More Details On \sys's Attention Design Implementation}
\begin{figure}[t]
\centering
\includegraphics[width=0.48\textwidth]{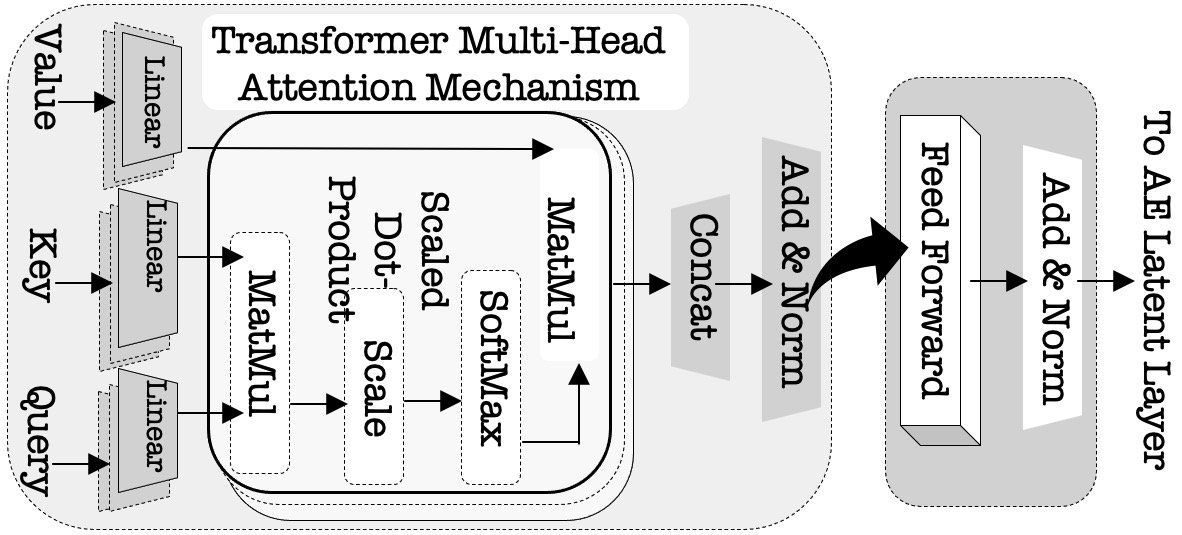}
  \caption{\sys's multi-head attention encoder and decoder to rank important Physics relationships and remove bottleneck of AEs}
\label{fig:multi-head}
\end{figure}

An attention function can be described as mapping a query and a set of key-value pairs to an output, where the query, keys, values, and output are all vectors~\cite{transformer}. The output is computed as a weighted sum of the values, where the weight assigned to each value is computed by a \emph{compatibility} function of the query with the corresponding key (shown in \fref{fig:multi-head}). \sys uses the scaled dot-product attention developed in \cite{transformer}, whose input consist of queries and keys of dimension $d_k$, and values of dimension $d_v$. In \sys, the query, key, and value matrices are simply copies of the input sequences, as shown in \fref{fig:multi-head}. Each is passed through different linear layers (layers with no activation), simply to map the input to the output and reduce dimensions and computational cost. The results of the query and key after being passed through the input layer are first multiplied (using dot product), scaled by  dividing each by the square root of $d_k$, and then passed through the \emph{softmax} function. The output is called the \emph{attention filter}. The attention filter is multiplied by the value matrix to obtain the weights on the values. The attention filter allows us to identify the important time series, i.e., actuator input and sensor outputs that have the most causal relationship. \sys uses the attention function to compute a \emph{rank} for each input in the sequence by applying the softmax function. In practice, \sys compute the attention function on a set of queries simultaneously in the matrix Q. K and V are matrices of the keys and values. 

\begingroup
\begin{equation} \label{eq:18}
Rank(Q,K,V) = SoftMax(\frac{QK^T}{\sqrt{d_k}})
\end{equation}
\endgroup
$PhysicsRank$ is \sys's attention function that ranks actuator sensor vectors with the most Physics causal relationship.

\PP{Parallelizing Attention}
\sys uses multiple attention heads (depicted in \fref{fig:multi-head} to identify useful attention filters for the input sequence, as well as compute the $PhysicsRank$ in parallel. That is, instead of performing a single $PhysicsRank$ function with $d$ dimensional keys, values, and queries, they can be projected $h$ times (where h is the Inertia-informed sequence size) to different learned $d_k$, $d_q$ and $d_v$ dimensions, respectively. The attention function, $PhysicsRank$, can now be performed on each of these projected versions of queries, keys, and values, to yield output values in $dv$ dimension. \sys then concatenates the results of the multiplication with each attention filter and then applies the linear layer to reduce (shrink down) the dimensions, and output the final values as shown in \fref{fig:multi-head}.

Internally, \sys implements the architecture of stacked self-attention and fully connected layers. The encoder and decoder process their input $x$ using a stack of identical layers equal to the ITB, each having two sub-layers, $Sublayer(x)$. The first is a multi-head self-attention mechanism and the second is a fully connected feed-forward network. \sys uses a residual connection followed by normalization around each sub-layers. We detail our implementation approach in the appendix in \fref{fig:multi-head}. The output of each sublayer in \sys is given below. Note that $Sublayer(x)$ is the function implemented by the sub-layer.
\begingroup
\begin{equation} \label{eq:17}
output(x) = LayerNorm(x + Sublayer(x))
\end{equation}
\endgroup
By using attention to prioritize causally-related actuator/sensor vectors within the ITB time window, \sys captures the periodic nature of a process in its learning algorithm.

\begin{figure*}[h]
\minipage{0.33\textwidth}%
\begin{subfigure}[c]{\textwidth}
\centering
     \includegraphics[height=0.7\textwidth]{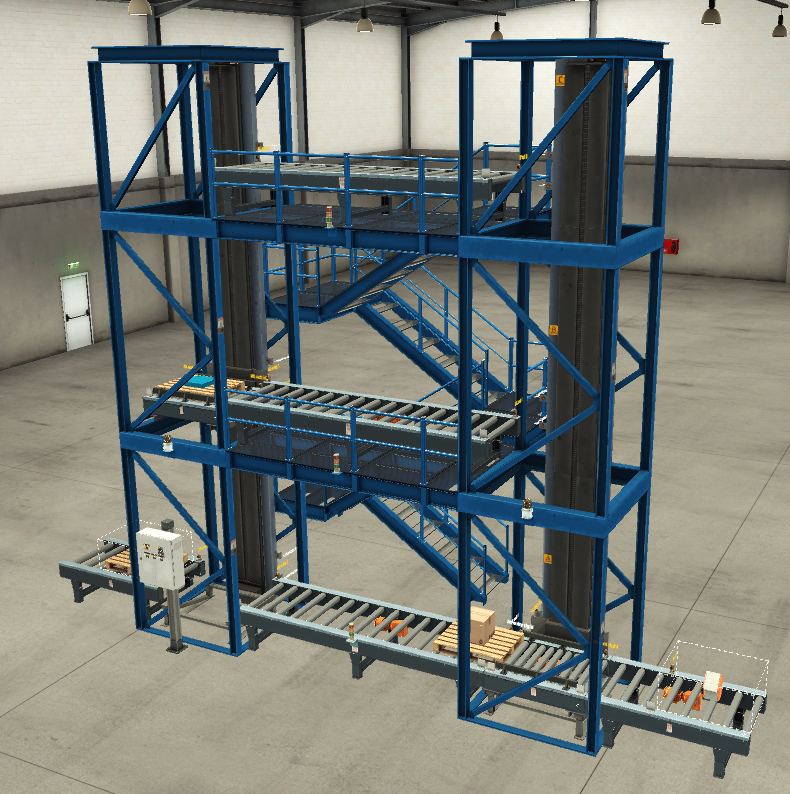}
    \caption{HMI for Elevator (Basic) Scene}
\label{fig:ElevatorAdvanced}
\includegraphics[width=0.7\textwidth]{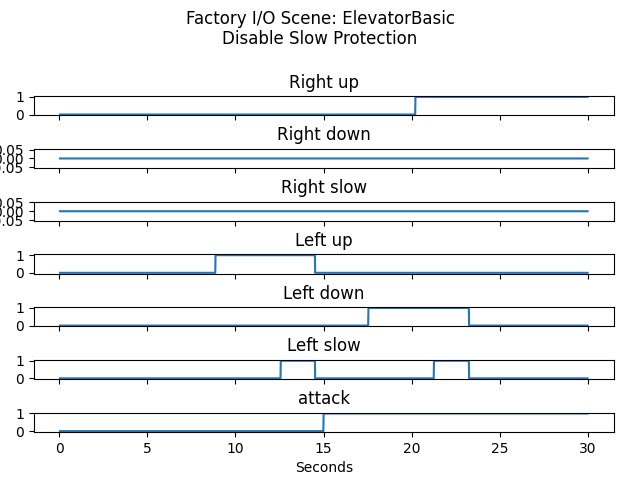}
    \caption{Graph of Elevator (Basic) Scene under attack}
\label{fig:ElevatorAdvancedGraph}
  	\includegraphics[width=0.78\textwidth]{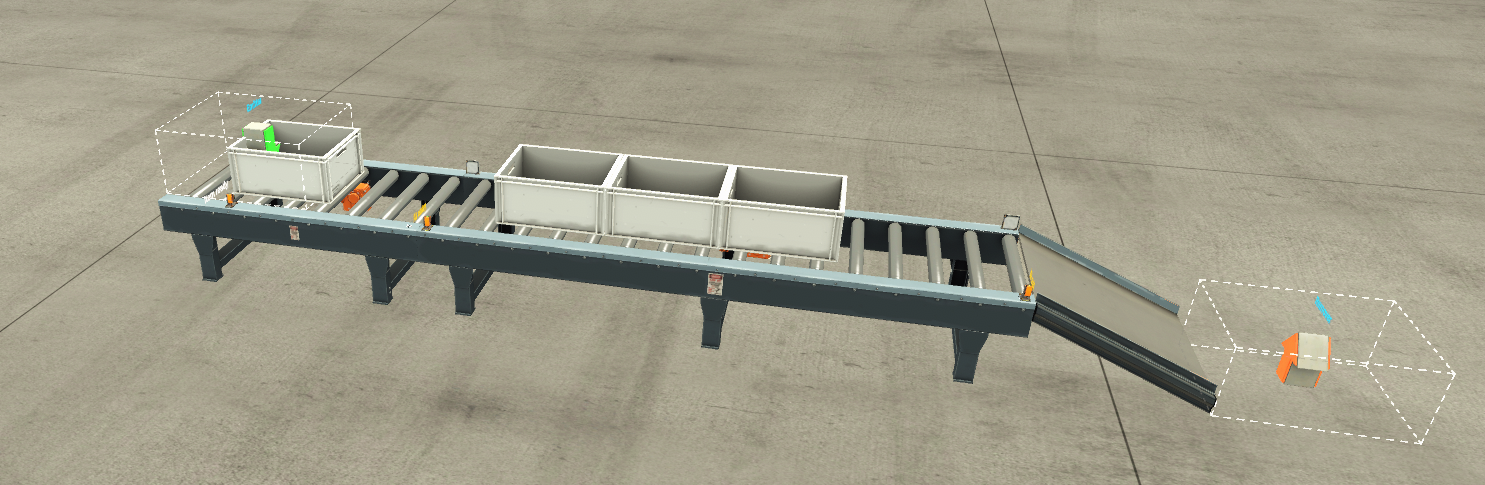}
    \caption{HMI for Queue Processor Scene}
\label{fig:QueueOfItems}
\includegraphics[width=0.78\textwidth]{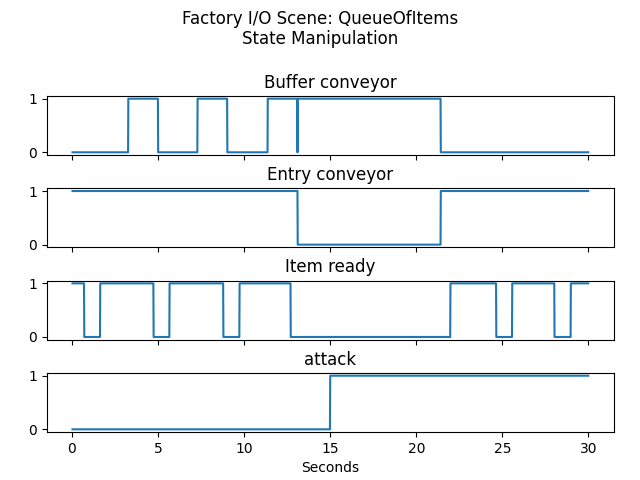}
    \caption{Graph of Queue of Items Scene under attack}
\label{fig:QueueOfItemsGraph}
\includegraphics[width=0.78\textwidth]{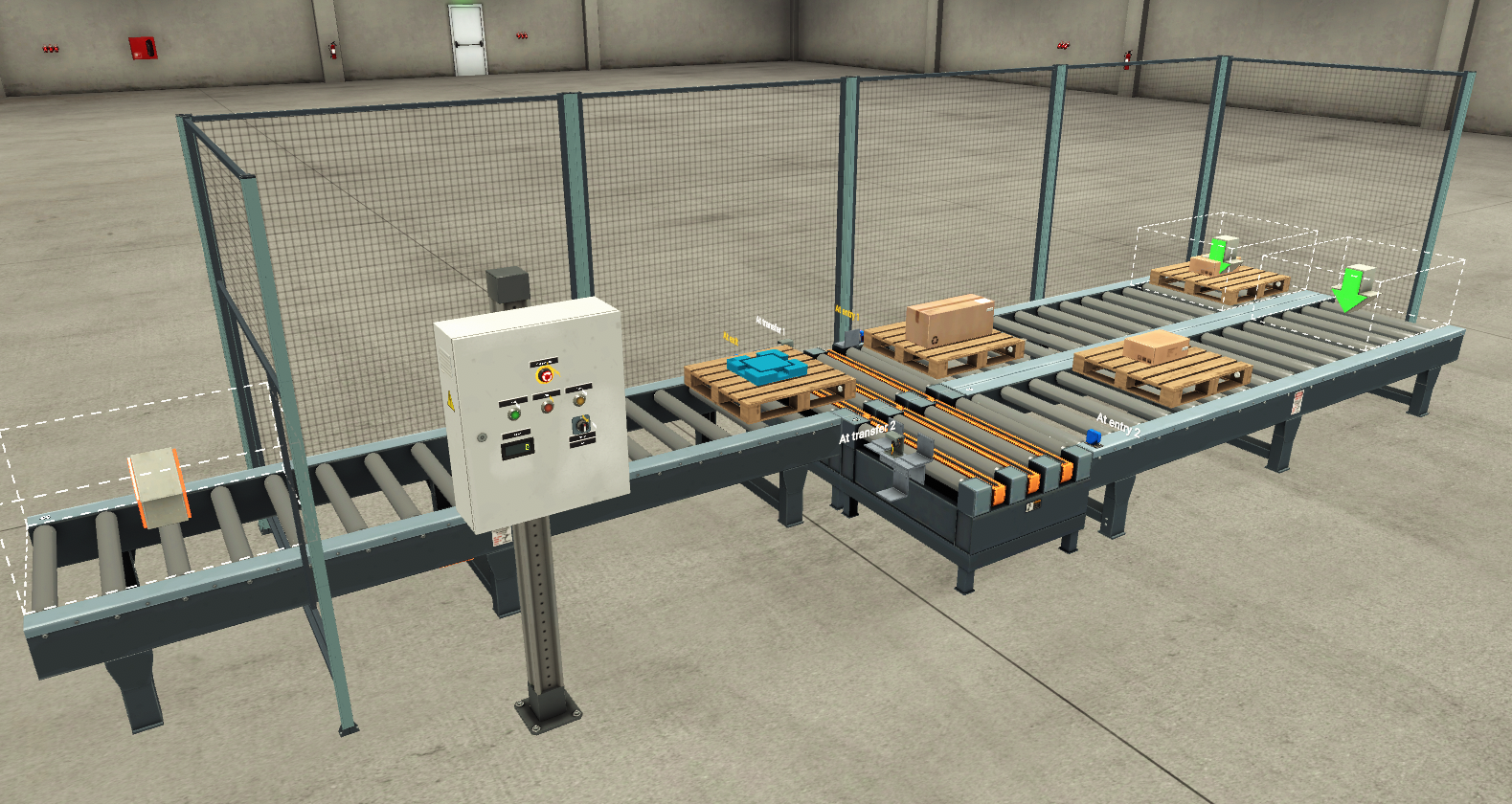}
    \caption{HMI for Converge Station Scene}
\label{fig:ConvergeStation}
\includegraphics[width=0.78\textwidth]{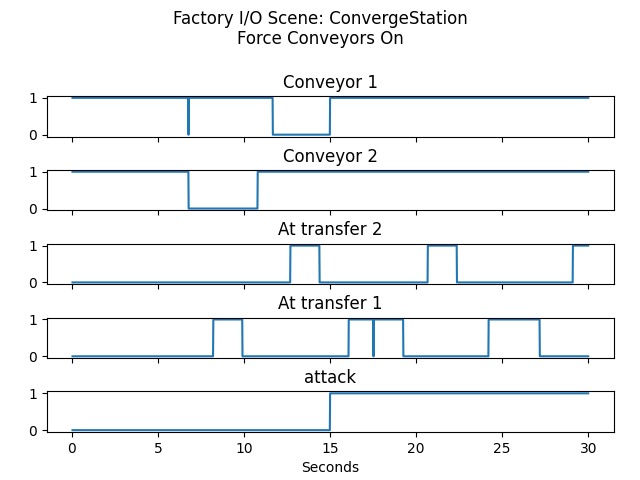}
    \caption{Graph of Converge Station's tags while under attack}
\label{fig:ConvergeStationGraph}
 \end{subfigure}
\endminipage
\minipage{0.33\textwidth}%
\begin{subfigure}[c]{\textwidth}
\centering
\includegraphics[width=0.78\textwidth]{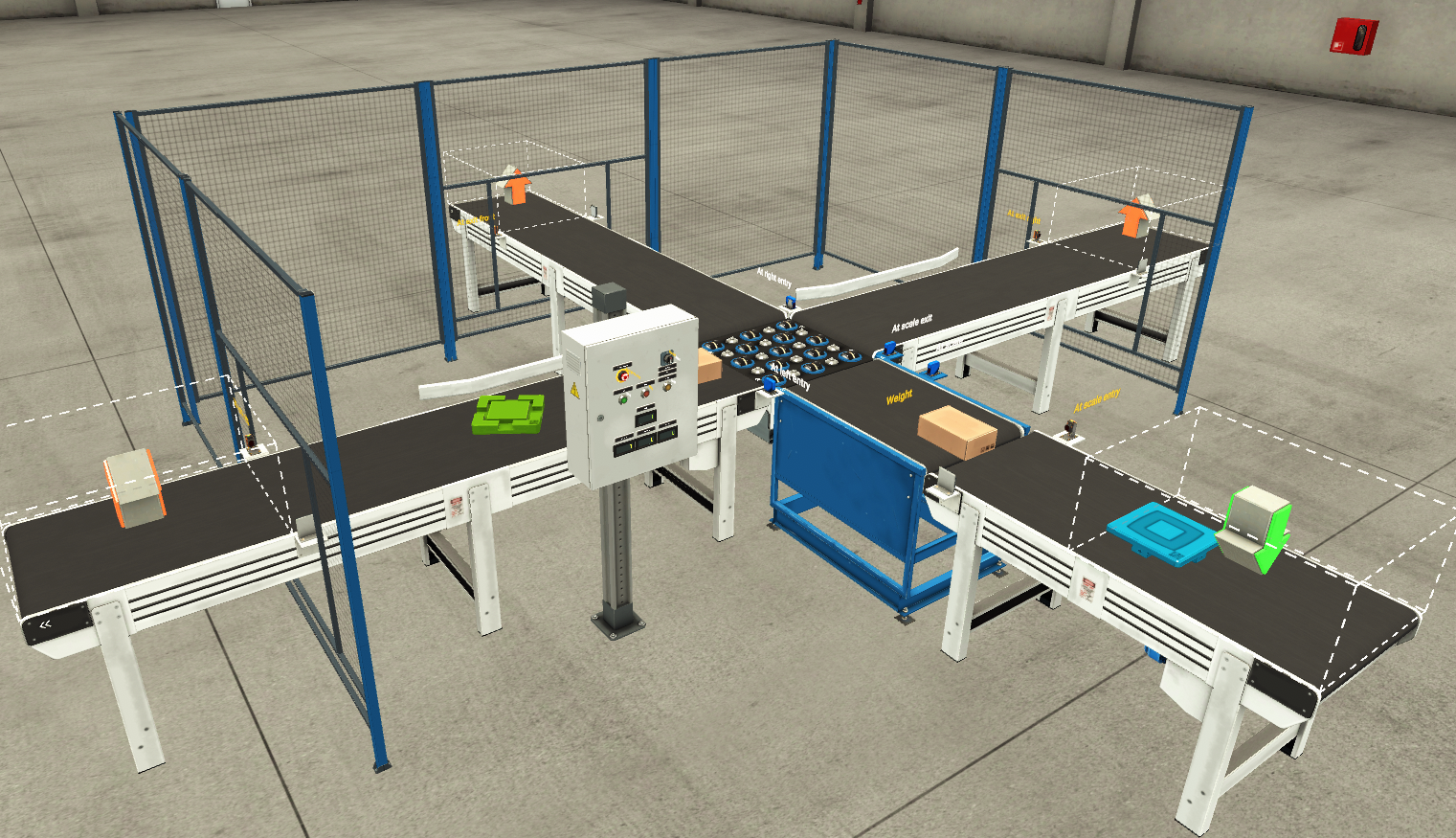}
    \caption{HMI for Sort By Weight Scene}
\label{fig:SortingByWeight}
\includegraphics[width=0.6\textwidth]{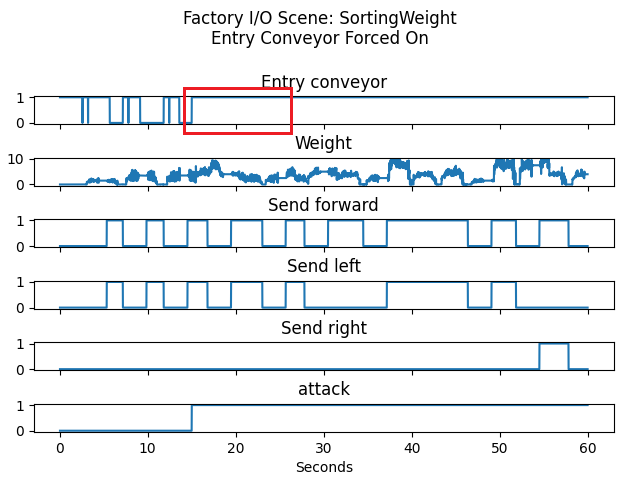}
    \caption{Graph of Sort By Weight's tags while under attack}
\label{fig:SortingByWeightGraph}
     	\includegraphics[width=0.6\textwidth]{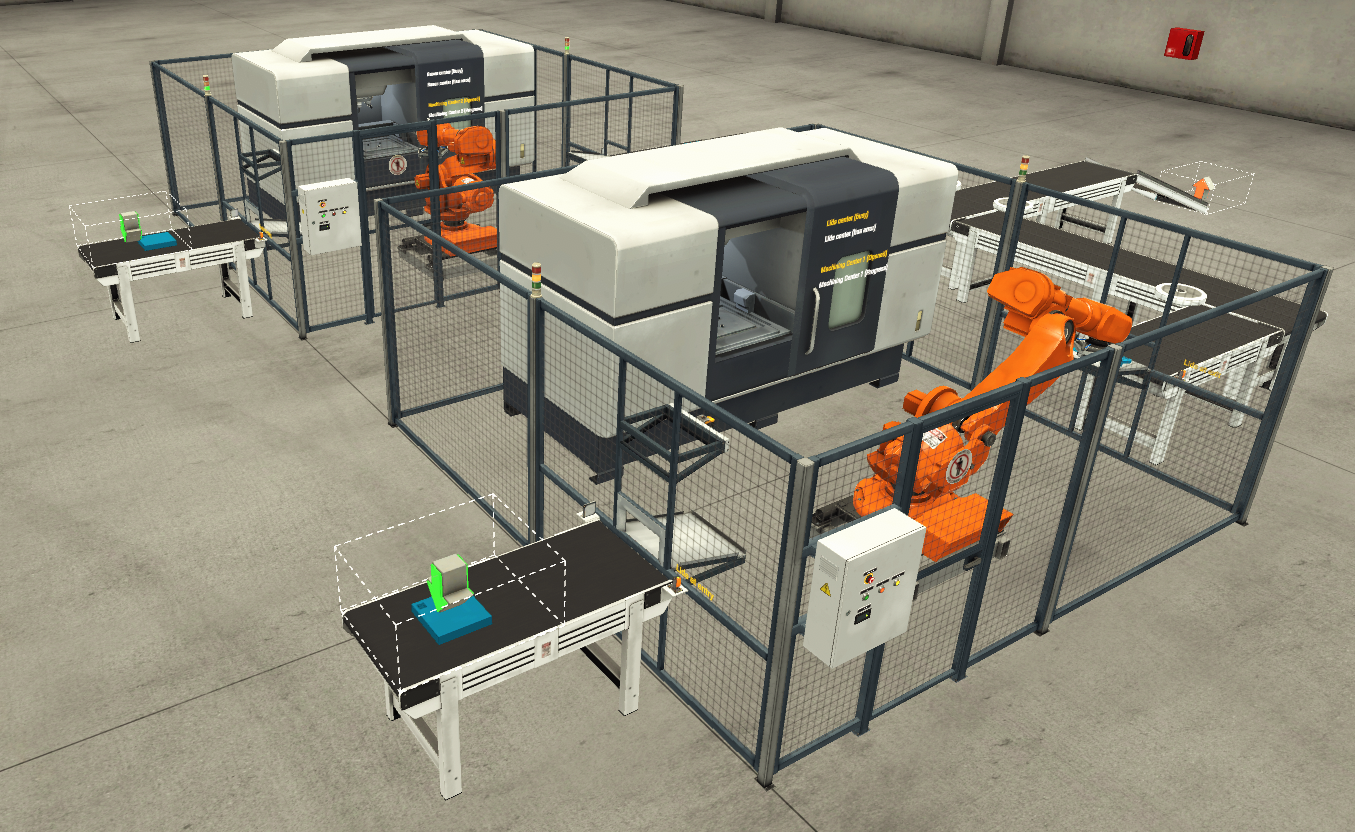}
    \caption{HMI for Production Line Scene}
\label{fig:ProductionLine}
\includegraphics[width=0.6\textwidth]{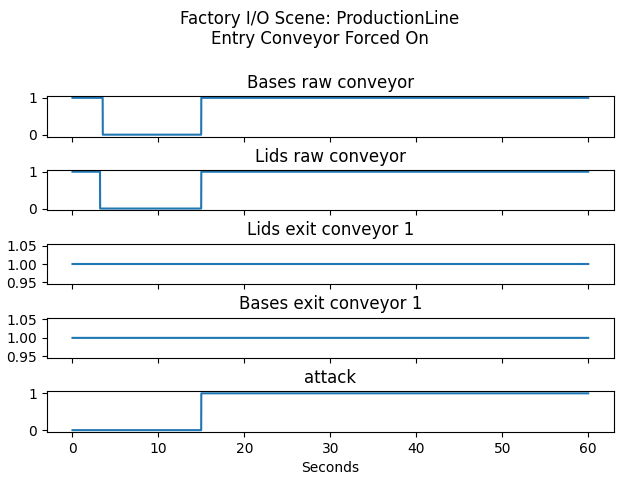}
    \caption{Graph of Production Line's tags while under attack}
\label{fig:ProductionLineGraph}
  	\includegraphics[width=0.6\textwidth]{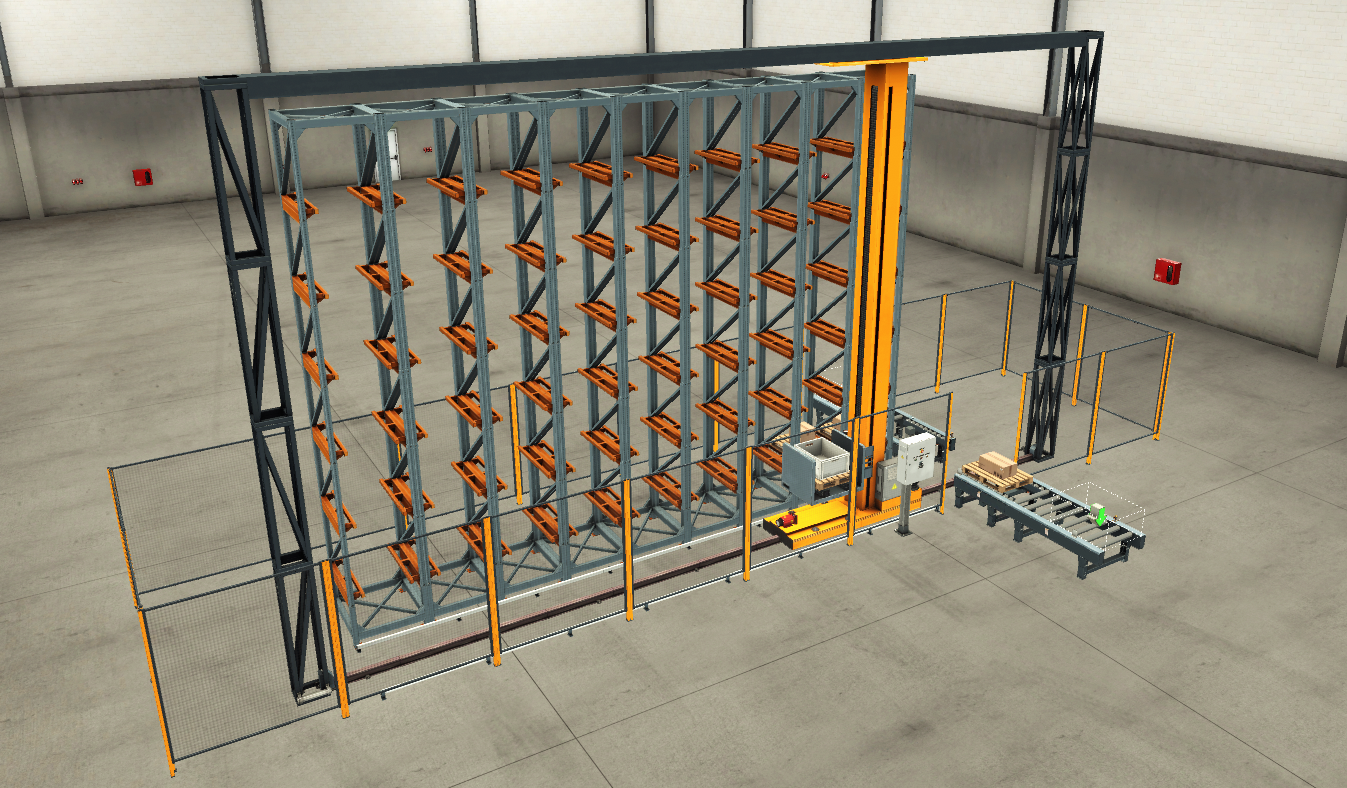}
    \caption{HMI for Automated Warehouse Scene}
\label{fig:automatedWarehouse}
\includegraphics[width=0.6\textwidth]{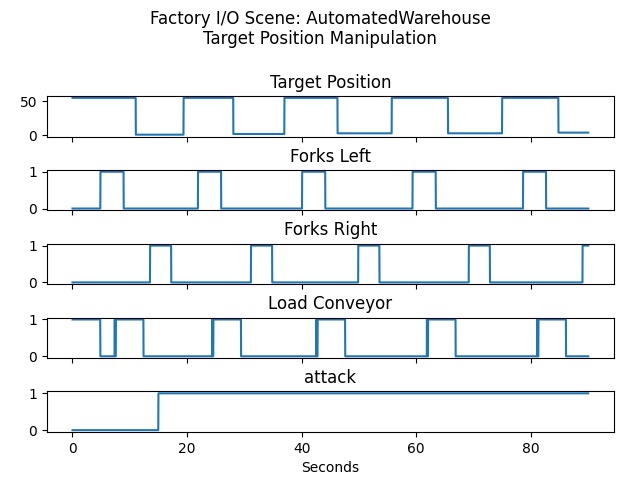}
    \caption{Graph of Automated Warehouse's tags while under attack}
\label{fig:automatedWarehouseGraph}
 \end{subfigure}
\endminipage
\minipage{0.34\textwidth}%
\begin{subfigure}[c]{\textwidth}
\centering
	\includegraphics[width=0.6\textwidth]{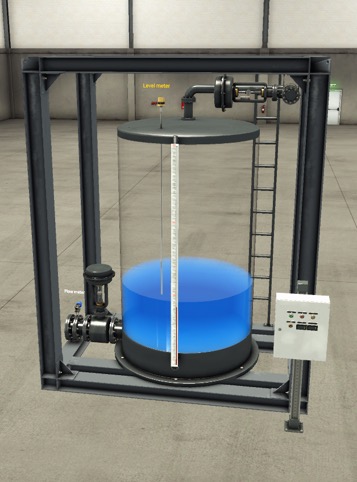}
    \caption{HMI for Chemical Dosing Scene}
\label{fig:LevelControl}
\includegraphics[width=0.6\textwidth]{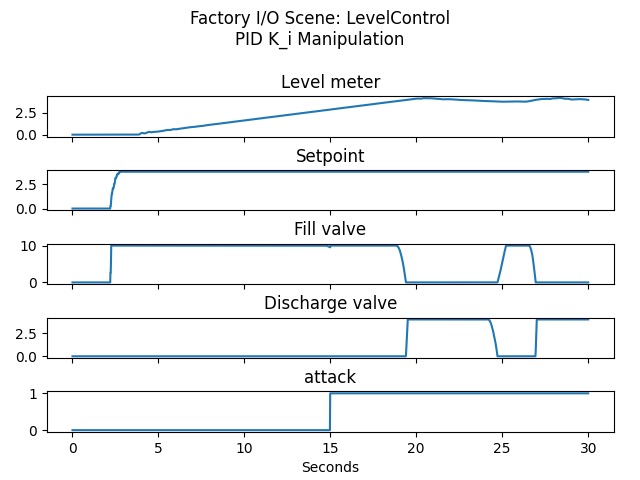}
    \caption{Graph of Level Control's tags while under attack}
\label{fig:LevelControlGraph}
     	\includegraphics[width=0.6\textwidth]{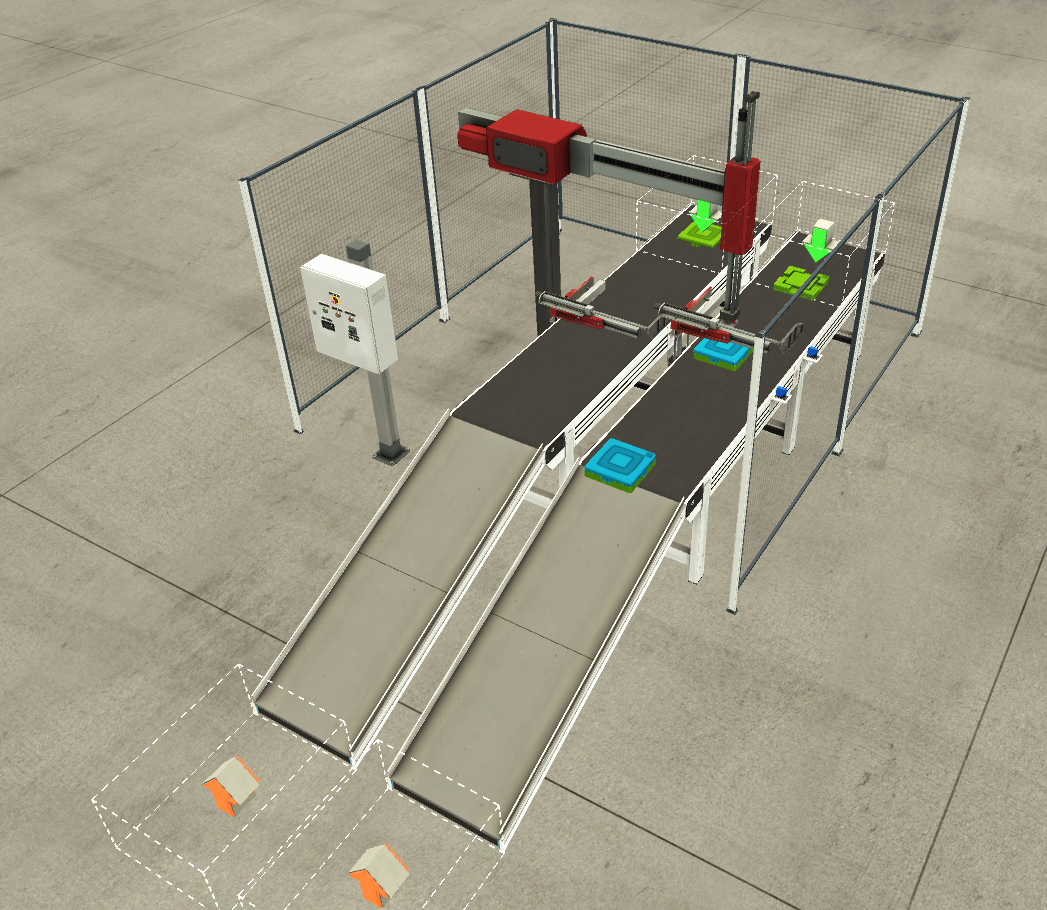}
    \caption{HMI for Assembler Scene}
    \label{fig:LevelControlGraph}
     	\includegraphics[width=0.6\textwidth]{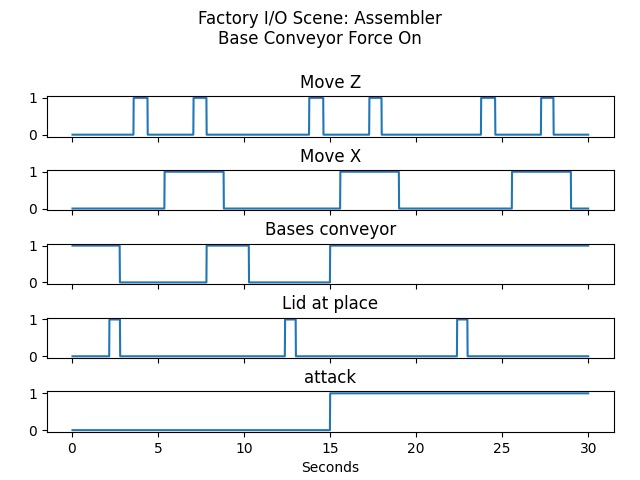}
    \caption{Graph of Assembler's tags while under attack}
\label{fig:AssemblerGraph}
  	\includegraphics[width=0.6\textwidth]{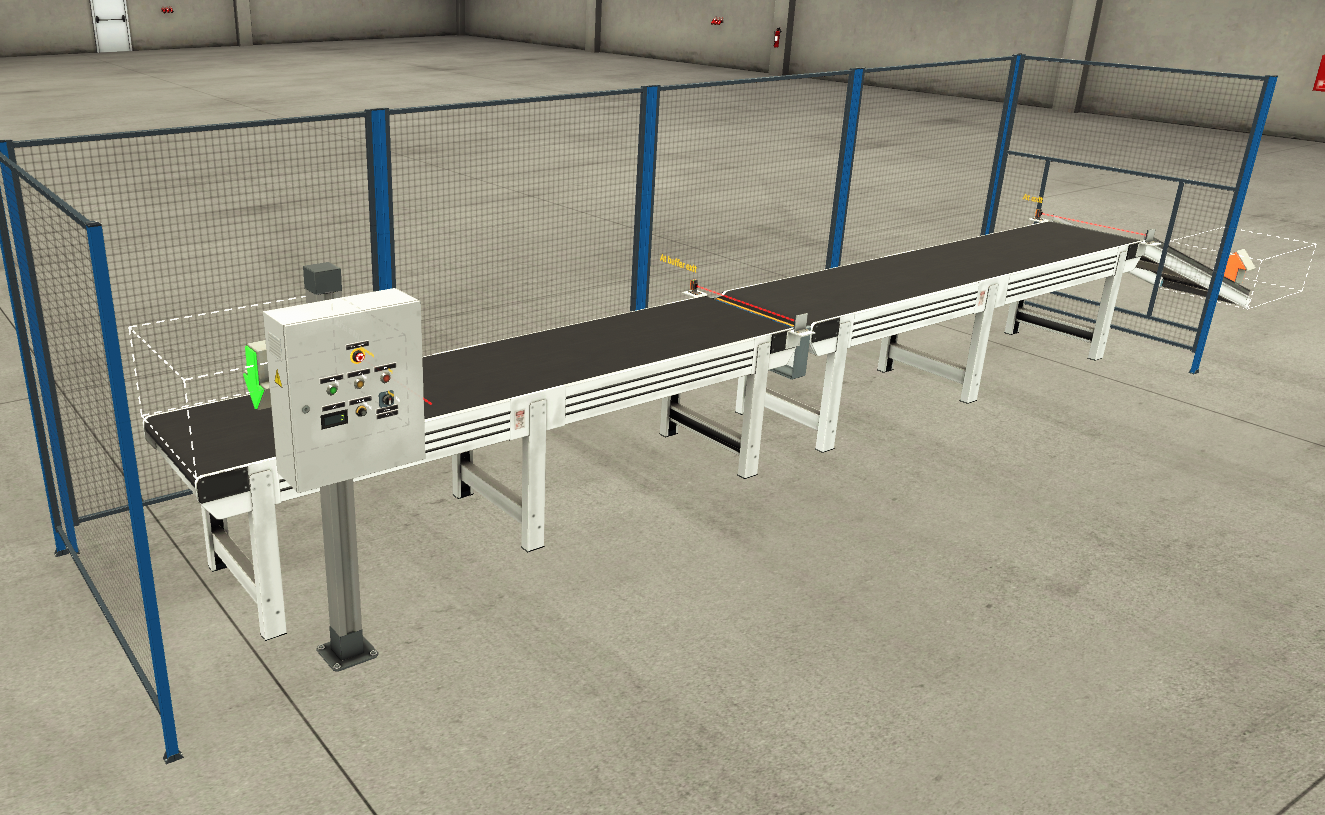}
    \caption{HMI for Buffer Station Scene}
\label{fig:BufferStation}
\includegraphics[width=0.6\textwidth]{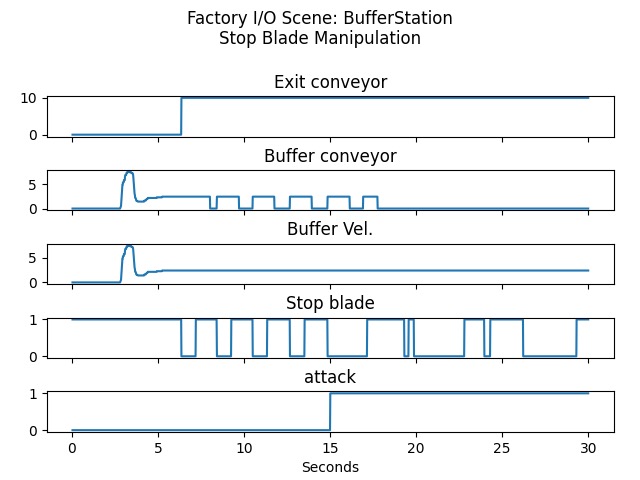}
    \caption{Graph of Buffer Station's tags while under attack}
\label{fig:BufferStation}

\label{fig:sortStationSymbolTableSPS}
 \end{subfigure}
\endminipage
\end{figure*}

\begin{figure*}[t]
\minipage{0.33\textwidth}%
\begin{subfigure}[c]{\textwidth}
\centering
\includegraphics[width=0.54\textwidth]{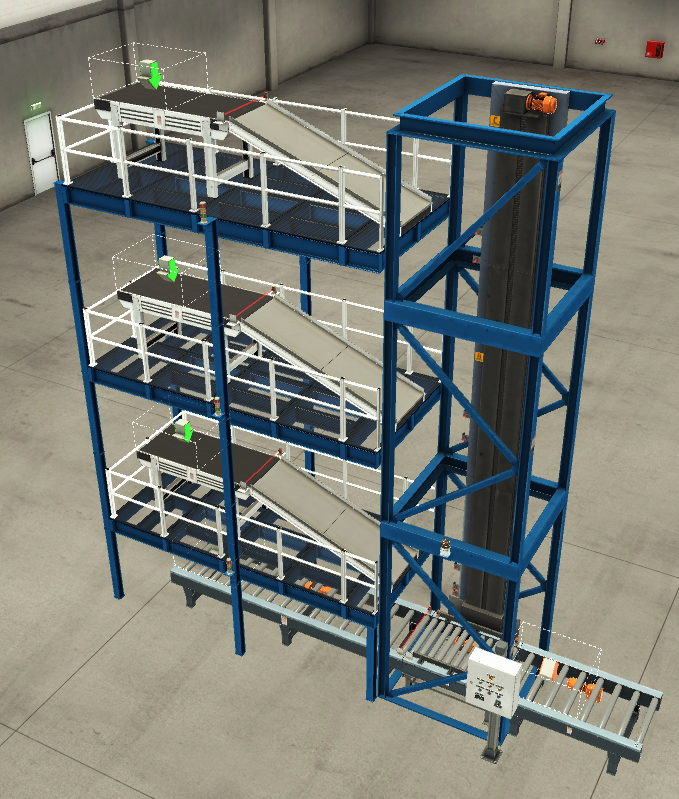}
\caption{HMI for Elevator (Advanced) Scene}
\label{fig:Elevator}
\includegraphics[width=0.7\textwidth]{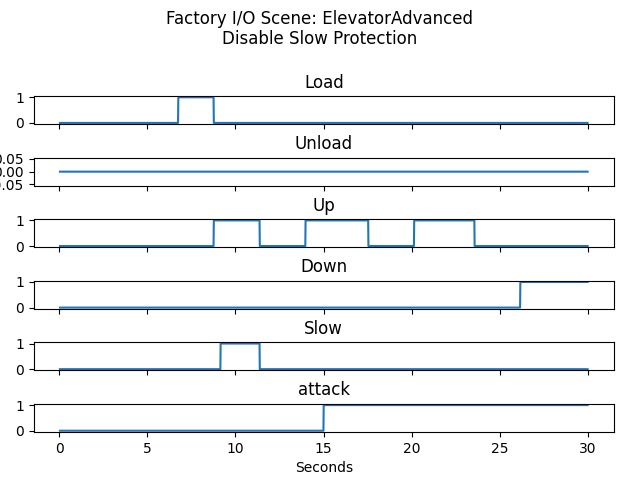}
\caption{Graph of Elevator (Advanced) under attack}
\label{ElevatorGraph}
 \end{subfigure}
\endminipage
\minipage{0.33\textwidth}%
\begin{subfigure}[c]{\textwidth}
\centering

\includegraphics[width=0.85\textwidth]{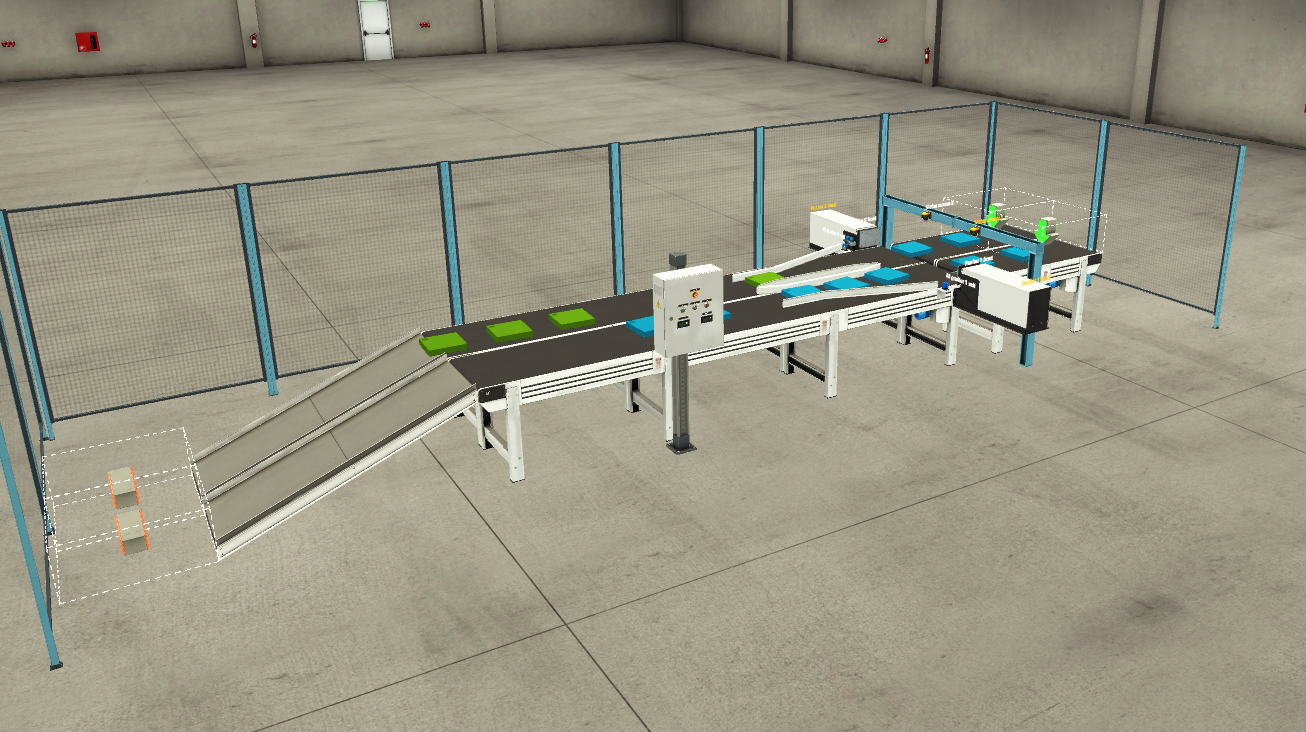}
    \caption{HMI for Separating Station Scene}
\label{fig:SeparatingStation}
\includegraphics[width=0.85\textwidth]{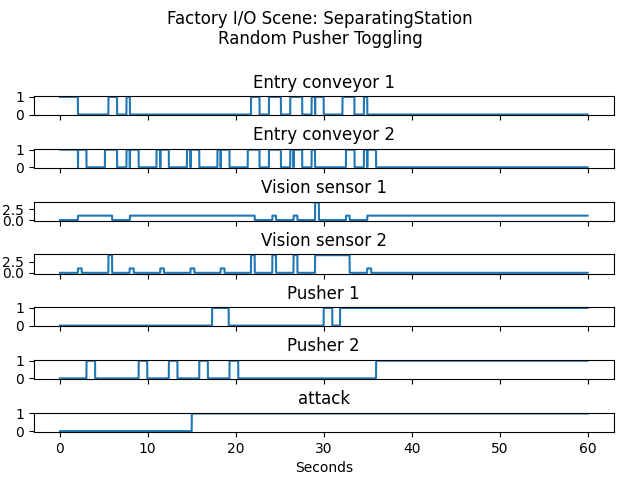}
    \caption{Graph of Separating Station's tags while under attack}
\label{fig:SeparatingStationGraph}
 \end{subfigure}
\endminipage
\minipage{0.34\textwidth}%
\begin{subfigure}[c]{\textwidth}
\centering
\includegraphics[width=0.85\textwidth]{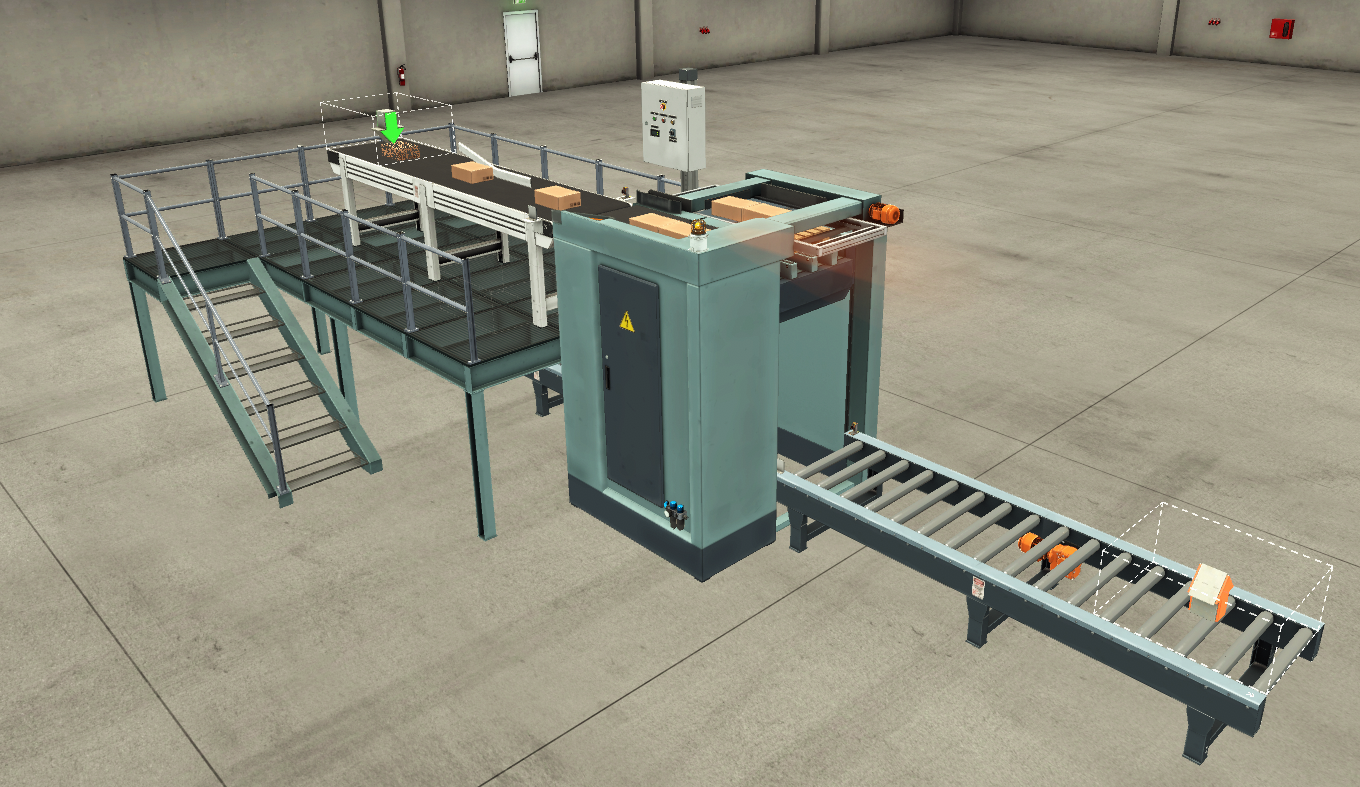}
    \caption{HMI for Palletizer Scene}
    \label{fig:ElevatorAdvanced}
    
    \includegraphics[width=0.85\textwidth]{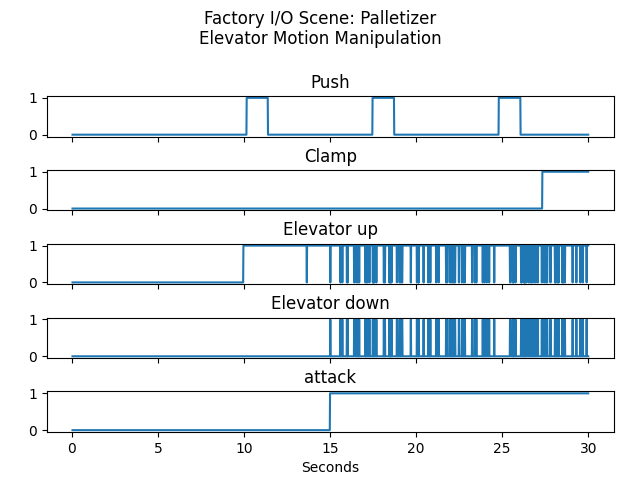}
    \caption{Graph of Palletizer's tags while under attack}
\label{fig:PalletizerGraph}
\label{fig:sortStationSymbolTableSPS}
 \end{subfigure}
\endminipage
\end{figure*}

\begin{figure*}[t]
\minipage{0.49\textwidth}%
\begin{subfigure}[c]{\textwidth}
\centering
	\includegraphics[width=1\textwidth]{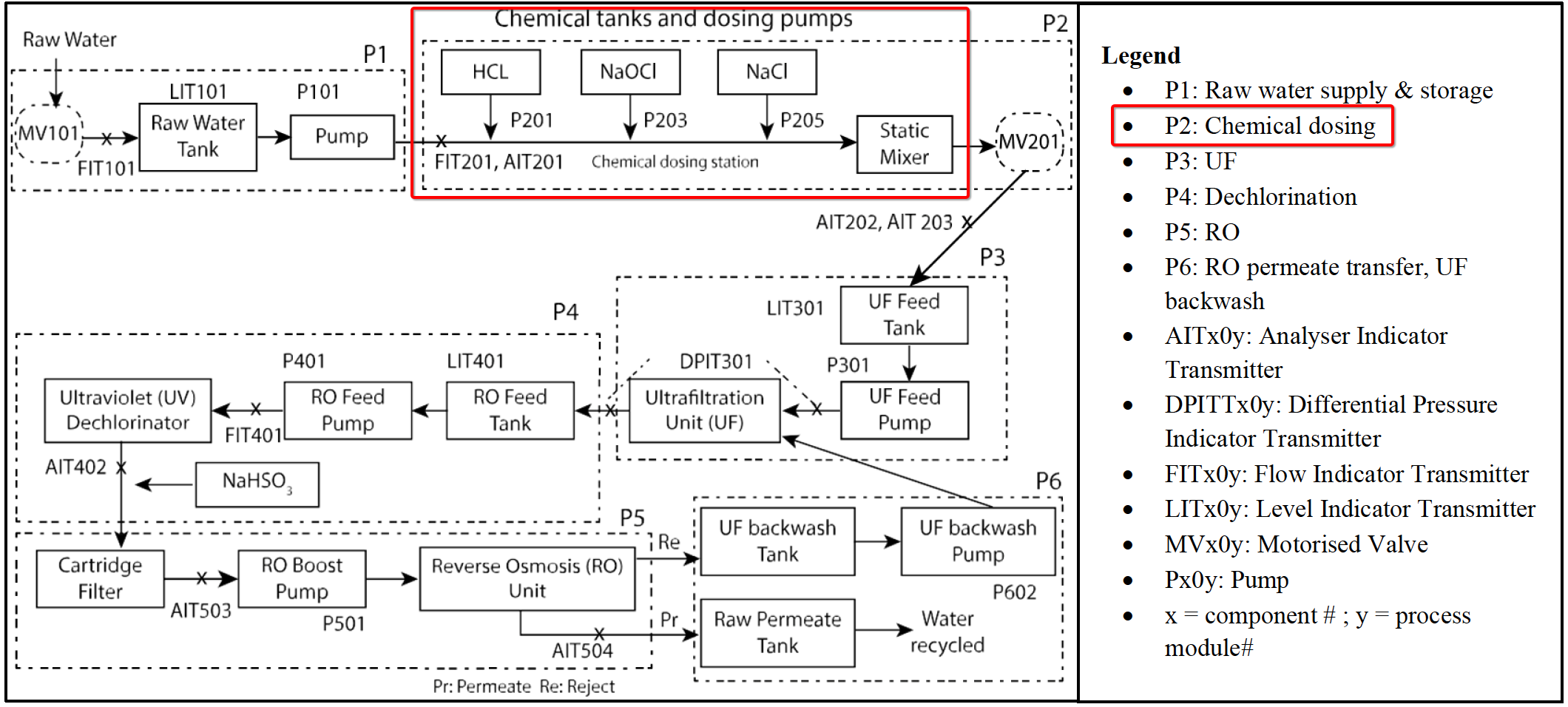}
    \caption{Complete water treatment plant based on~\cite{swat1, swat2}: The chemical dosing operation was attacked in the Oldsmar water poisoning attack}
\label{fig:wt}
 \end{subfigure}
\endminipage \hspace{1em}
\end{figure*}

\end{document}